# Who Falls for SMiSh? Learning Through Survey Data Where to Best Target Awareness Training for Mobile Messaging Attacks

Cori Faklaris, Sarah Tabassum, and Heather Richter Lipford, *UNC Charlotte*

## Abstract

As mobile phone adoption has surged, so have scams involving these devices. One such scam, known as "SMiShing" (or "smishing") after Short Message Service (SMS), involves fraudsters sending phishing links via mobile texts. Despite the prevalence of SMiShing, there is a lack of data on who is most vulnerable to these attacks. Prior research on phishing (its email counterpart) suggests that susceptibility may vary by demographic and contextual factors. In two large-scale surveys, we use a previously published simulation method to collect data from representative samples of U.S. adult mobile phone users. Our findings indicate that younger individuals and college students are particularly vulnerable. Participants struggled to correctly identify legitimate messages, with the second study providing comparisons of financial message variants. Researchers, regulators, and telecoms can help users by creating mobile-specific interventions for under-24 and university customers and adding verifications and warnings.

## 1. Introduction

As more people use mobile phones [44], scams targeting these devices have also increased [45]. By the end of 2022, the most common method scammers used to contact people in the U.S. was through phone calls and texts [42]. SMiShing, or "smishing" (after SMS or Short Message Service), is a type of scam where fraudsters send phishing links via text messages. Common targets include banks, delivery companies, retailers, and communication providers [46]. Despite the rise in SMiShing, we do not have enough data on who is most vulnerable to these attacks. Such data will help in designing more-nuanced interventions that avoid a one-size-fits-all approach, targeted to those who can most benefit from them.

Previous research on email phishing (e.g. [5, 11, 12, 34]) suggests that vulnerability to SMiShing might vary based on demographic and contextual factors. However, no study has systematically examined which specific factors make people more likely to fall for SMiShing. Understanding this is crucial for developing effective awareness training and interventions. For example, this data can help financial institutions to design risk communication and education programs for their account-holders, and telecom providers to improve notifications and warnings for mobile users.

To start addressing this gap, we conducted two large-scale online surveys of U.S. adult mobile phone users. With the resulting data, our paper answers the following questions:

- ***RQ1****: How many U.S. adult mobile phone users can correctly identify whether three random text messages are legitimate or fraudulent?*
- ***RQ2****: Which U.S. demographic groups are most vulnerable to SMiShing, based on survey responses?*
- ***RQ3****: How is the vulnerability identified in RQ1 associated with prior training or other relevant experiences?*

We designed an online assessment to see if people could tell whether a simulated text message was "real" or "fake" [30]. We collected various demographic, cognitive, and behavioral data, such as age, education, and security training. To answer RQ1, RQ2, and RQ3, we surveyed 1,007 U.S. mobile phone users, using a mix of real and SMiSh messages. We conducted a follow-up survey focusing on variants of a common SMiSh message, a Zelle bank payment notification.

In Study 1, participants were better at identifying fake messages (81.4%) than real ones (23.5%). Study 2 showed similar results, with 68.7% correctly identifying fake messages and 58.8% identifying real ones. We found that younger people and college students scored the worst on the online SMiShing assessment. Gender difference in vulnerability seen in previous studies did not appear in our results. We also found that people who frequently experienced security breaches scored lower on our SMiShing assessment, while those working in healthcare, education, office-related, and food-service job categories scored higher. In Study 2, younger people and college students significantly misidentified real messages when controlling for bank and Zelle usage. Results for message variants underscore that header information is not sufficiently helpful for judging scams.

Based on these findings, we recommend researching more-targeted risk-awareness campaigns and educational interventions, especially for young adults and college students. Our data suggests that new ideas are needed, as we found no significant effect of self-reported amounts of security awareness and SMiSh-relevant training. As a training backstop and additional layer of mitigation, we also recommend that U.S. regulators and telecom providers work with usability experts to improve verification and guidance for mobile messaging. This would help users to more easily identify trusted sources. Finally, we suggest more research to explore the relationships among age, educational attainment, security training, and SMiShing vulnerability for financial providers who are often impersonated by scammers.

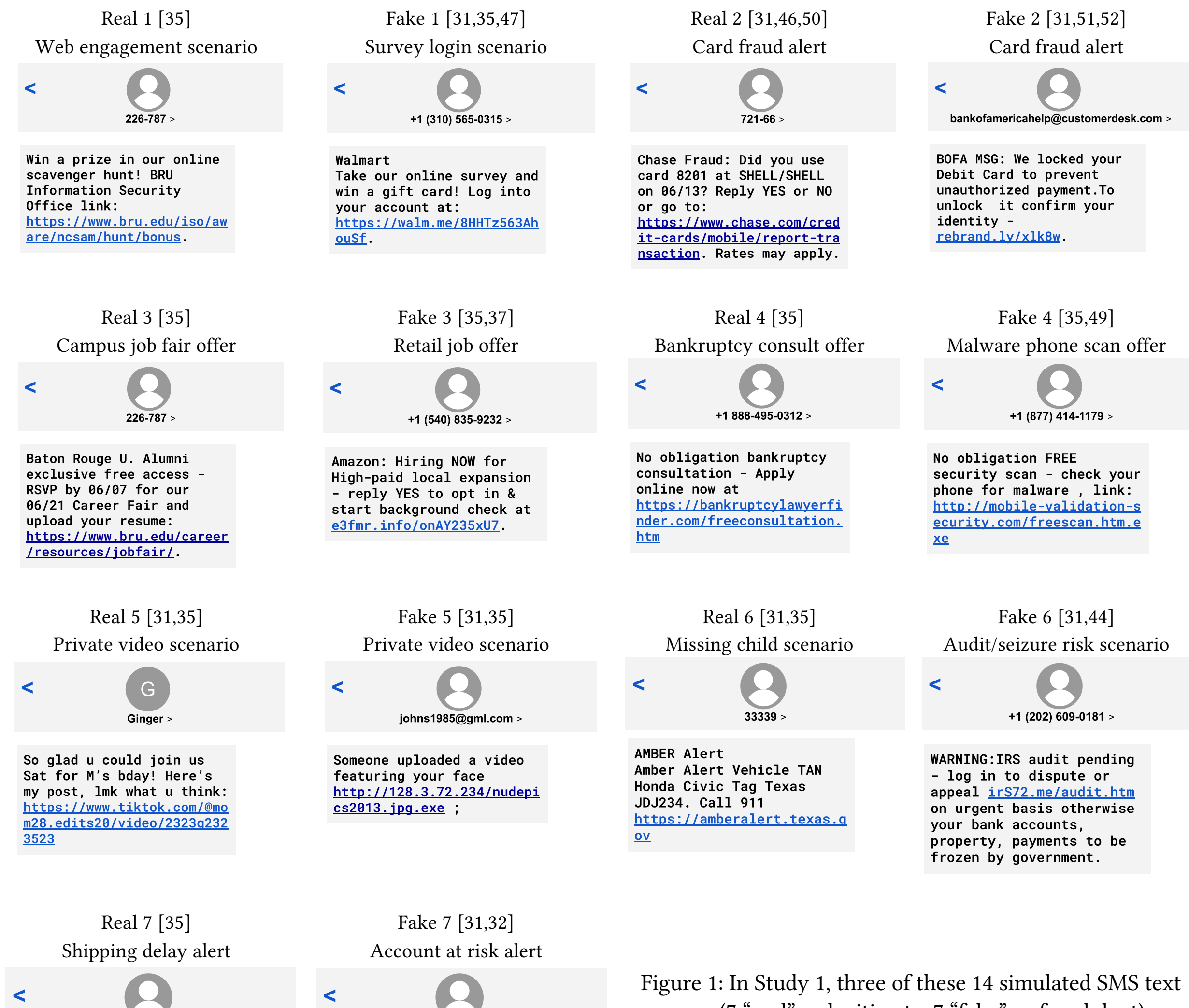

Figure 1: In Study 1, three of these 14 simulated SMS text messages (7 "real" or legitimate, 7 "fake" or fraudulent) were randomly served to each participant to rate and answer questions about. Each included a URL and other clues to help guide the participants' judgments, such as whether it used an SMS short code or other source identifiers and whether it included typos or odd grammar and spacing.

## 2. Background and Related Work

Phishing and SMiShing are cyberattacks that use social engineering to trick users into revealing personal or financial information via computer-mediated communication [18]. These attacks threaten user security and privacy and can harm organizational reputation.

While there is substantial data on demographic susceptibility to phishing, it is unclear how these findings apply to SMiShing. This study aims to bridge that gap by examining demographic and contextual factors associated with SMiShing vulnerability as found in large-scale and representative U.S. survey samples.

### 2.1 Phishing

Phishing is a common and well-studied cyberattack [47]. Attackers use fraudulent emails to impersonate legitimate entities and solicit sensitive information [22]. Phishing emails often contain malicious links or attachments that lead users to fake websites or download malware. These attacks can target individuals or organizations, aiming to steal money, identities, credentials, or intellectual property. Machine learning methods, user interfaces, and training have shown effectiveness in mitigating phishing [22].

Many studies have investigated factors influencing phishing susceptibility, such as email design, spelling, phrasing, URL formatting, brand spoofing, context, and user characteristics [3, 10–12, 14, 23, 26, 34, 38]. Sheng et al. [34] were among the first to investigate demographic vulnerability, finding that women and younger people were more susceptible to phishing. Educational materials reduced willingness to enter information into bogus webpages but also decreased users' tendency to click on legitimate links.

Like previous studies in phishing, our studies displayed simulated scam and real messages for participants to respond to, adapting some examples and survey items from prior work.

### 2.2 SMiShing

SMiShing, or "smishing," uses fraudulent SMS text messages to deceive users into clicking on malicious links, calling a number, or providing personal information. SMiShing messages often exploit emotions like fear, love, or greed to induce immediate action without verifying the source [46]. These attacks can leverage users' trust in services or brands, such as banks, delivery companies, or online retailers, or in authorities, such as security or military [43]. For example, a SMiShing attack on customers of the U.S. Fifth Third Bank led them to enter their credentials on a bogus website [40]. Another attack tricked Czech Post customers into downloading a malicious app [4]. A third attack exploited COVID-19 information confusion to send bogus messages from contact tracing websites, insurance, or vaccine providers. [2].

SMiShing attacks are harder to detect than email phishing because text messages have fewer cues, such as sender's address or subject line [48]. Text messages are also more likely to be read and responded to than emails, as they are perceived as more personal and urgent [7]. Mishra and Soni [25] point out that the use of mobile devices, with their small screens and typical lack of security awareness, makes users particularly vulnerable to smishing attacks.

Some researchers have systematically studied SMiShing vulnerability. Tabassum et al. interviewed 29 mobile users, discovering that they focus more on the content of messages than the sender information when judging whether the messages are fake [36]. Rahman et al. conducted a real-life experiment with 10,000 participants, finding that personalized or spoofed messages were more effective [30]. However, they did not mix legitimate SMS messages with fraudulent ones, and few participants answered their follow-up survey. Using the simulation survey method (n=187), Timko et al. found that participants who were shown 16 SMS screenshots had 67.1% accuracy with fake messages and only 43.6% with real ones [37].

Our Study 1 adapts elements of these studies, such as the types of SMiSh messages that are shown and what cues they include. Study 2 goes beyond Timko et al. to experimentally test “Real” and “Fake” variations on one message.

Table 1: For Study 1 participants, counts for demographics, tech usage, experience working with sensitive data, and jobs.

| Age | | Education | | Household Inc. | | Gender Identity | | Hisp./Lat./Sp.? | | Racial/Ethnic Identity | | Household Size | |
|---|---|---|---|---|---|---|---|---|---|---|---|---|---|
| 18-24 | 232 | No 4y deg. and not in school | 537 | < $26.5K poverty line | 188 | Female | 630 | No | 900 | White/Cauc. | 839 | 1 ppl. | 175 |
| 25-34 | 70 | No 4y deg., but in school | 129 | $26.5-$49K | 171 | Male | 362 | Yes | 96 | Black/African | 106 | 2 ppl. | 320 |
| 35-54 | 312 | 4y deg., but no doctorate | 232 | $50-$99K | 358 | Nonbinary | 10 | Prefer not to say | 10 | Asian – total for all regions | 24 | 3 ppl | 160 |
| 55-64 | 173 | Has doctorate | 108 | $100K+ | 289 | Self-described | 2 | | | Native Am. or Alaska Native | 9 | 4 ppl. | 226 |
| 65+ | 219 | | | | | Prefer not to say | 2 | | | Self-described | 15 | 5+ ppl. | 125 |
| | | | | | | | | | | Prefer not to say | 13 | | |

| Primary Mobile Phone | | Usage / Past Week | | Exp. w/Sens. Data | | Primary Job Status | | Top Occupations (FT or PT) | |
|---|---|---|---|---|---|---|---|---|---|
| Smartphone - Android | 531 | <6 hrs. | 167 | None at all | 390 | Full-time (FT) | 433 | Sales and Related | 51 |
| Smartphone - iOS/Apple | 452 | 6-10 hrs. | 208 | Only a little | 149 | Part-time (PT) | 112 | Business/Financial Ops. | 46 |
| Other smartphone | 7 | 11-20 hrs. | 248 | A moderate amount | 191 | Unemployed | 148 | Computer/Mathematical | 44 |
| Featurephone with camera | 11 | 21-30 hrs. | 190 | A lot | 137 | Retired | 134 | Office/Admin. Support | 44 |
| Basic phone with no camera | 5 | >30 hrs. | 193 | A great deal | 139 | At-home parent | 54 | Construction and Extraction | 38 |
| | | | | | | Self- employed | 42 | Healthcare Practitioner and Technical | 38 |
| | | | | | | Unable to work due to disability | 27 | Educational Instruction and Library | 37 |
| | | | | | | FT student | 43 | Food Prep. and Service | 36 |
| | | | | | | PT student | 13 | Management | 36 |

### 2.3 Demographic Susceptibility

Prior work has found gender and age to be significant demographic factors influencing susceptibility to both email and SMS phishing, although specific vulnerabilities vary. Sheng et al. and Yang et al. [34, 39] found that women were generally more susceptible than men. Lin et al. [24] found an interaction between age and gender in email phishing, noting that older women are the most vulnerable group. Regarding age, Sheng et al. [34] found that those aged 18-25 are particularly vulnerable to email phishing, while Abroshan et al. [1] found no association between age and phishability, though they did find that women were more prone to clicking on a phishing link. Yang et al. [39] found that age was an important factor in falling for phishing emails, along with education and personality.

For SMiShing, Rahman et al. [30] found that study participants aged 45-54 were more susceptible to the smishing messages that they were sent, while the 18-24 age group was the least susceptible. This contrasts with many email phishing findings, where younger adults were often more vulnerable. Rahman et al. also found that participants with a doctoral degree fell for the simulated smishing more than those with less education.

Our work uses representative survey samples to collect demographics such as gender, age, and educational attainment, including whether participants had a doctorate, had already earned a four-year college degree, are currently studying for a four-year degree, or have no four-year college experience. This enables comparisons with the prior demographic results on phishing and SMiShing.

## 3. Methods

We designed and deployed an online assessment to collect data on demographic vulnerability to SMiShing and related factors. This method allows us to investigate the factors contributing to susceptibility to SMiShing.

### 3.1 SMiShing Assessment Design

For Study 1, we created 14 simulated text messages, half legitimate and half fraudulent, to test participants' ability to identify SMiSh messages. This method avoids ethical issues like panicking users [6, 13, 34]. We adapted examples from prior work and real-life SMS messages, incorporating entities like "IRS," "Walmart," and "Facebook" [30]. Each message included a URL, a reward or fear motivator, and credibility clues [21]. Many incorporate cues that Tabassum et al. [36] found mobile users rely on to spot the SMiSh, such as URL and call-to-action elements (perceived as scams) and official-looking format (perceived as legitimate). For the scoring, participants rated three randomly served messages from this set as: 1=Fraudulent, 2=Likely Fraudulent, 3=Not Sure, 4=Likely Legitimate, 5=Legitimate. Because we have ground-truth labeling of each message as either legitimate or fraudulent, we reasoned that we could translate the participants' ratings into a second, binary variable representing whether their judgment was correct (Section 3.5), without adding a second item and lengthening the survey further. See Figure 1 for the images and Appendix A for the text.

Next, we piloted our questionnaire with in-person cognitive or "think aloud" interviews [28, 41] (*n*=2), reviews with our study team and fraud researchers in industry (*n*=6) and remote surveys on Prolific (*n*=11). These showed that the questionnaire could be completed within about 12 minutes. A longer survey is undesirable because participants may become fatigued and quit [29, 32].

One feedback in early pilots was that knowing whether someone has an account with the entity in the SMS text message helps them judge whether it is fraudulent or legitimate. We tested in the main study whether it affects answers to have participants answer as "yourself" or as "Pat Jones," hypothetical owner of many accounts [34]. No significant effect was found, and we do not include this covariate in reporting Study 1 results. See Appendix B for details.

For Study 2, we zeroed in on comparing "real" and "fake" variations of a Zelle transaction notification from Bank of America, a common SMiSh vector [9, 49–51]. To confirm that the questionnaire remained easy to use and understandable, we reviewed the messages within the study team (*n*=5) and piloted the survey on Prolific (*n*=3). Participants rated messages as 1=Real or 2=A scam. We also asked participants whether they were a Zelle user or Bank of America customer, which allowed us to test their effect on the message ratings. The messages are reproduced in Figure 5 and Appendix A.

### 3.2 Covariate Data Collection

For Study 1, Qualtrics recruited a panel of 1,000 U.S. mobile phone users aged 18 or older, matching U.S. Census data for age, income, and education [52]: 18-34: 30%, 35-54: 32%, 55+: 38%; <$50K: ~35%, $50K-100K: ~35%, 100K+: ~30%; no college degree: 65%, 4-year degree or higher: 35%. Participants meeting these quotas were asked detailed questions about age, income, education, gender, racial/ethnic identities, household size, experience with handling sensitive data, and occupation status and job category, per the U.S. Bureau of Labor Statistics. At the end, participants completed an attention check and items assessing internet and information-security experiences, attitudes, behavior intentions, and prior training on responding to phishing and SMiShing. These variables, collected in prior mental-models studies (such as those by Redmiles et al. [31, 33], Faklaris et al. [20, 35] and Egelman et al. [15–17]) provide a statistical snapshot of the sample. These demographic variables are crucial for answering RQ2 and RQ3.

Along with demographics as identifiers, the questionnaire collected IP addresses and device metadata, to enable us to map responses and to test for effects from device modality and operating system. We did not collect other identifiers, to encourage free responses and because the data was not needed to answer the research questions. The Institutional Review Board approved this as an exempt study under Category 2 of the U.S. Revised Common Rule.

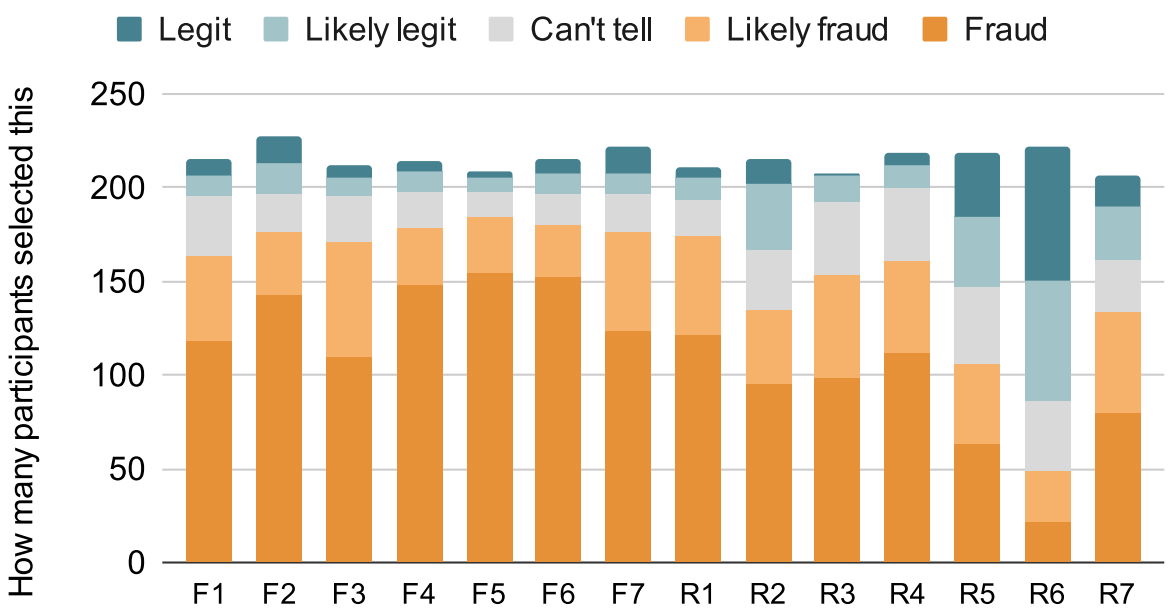

Figure 2: In Study 1, a majority of participants correctly rated all 7 simulated "fake" SMS text messages as Likely Fraudulent or Fraudulent. For 5 of 7 simulated "real" text messages, a majority incorrectly rated them as Likely Fraudulent or Fraudulent.

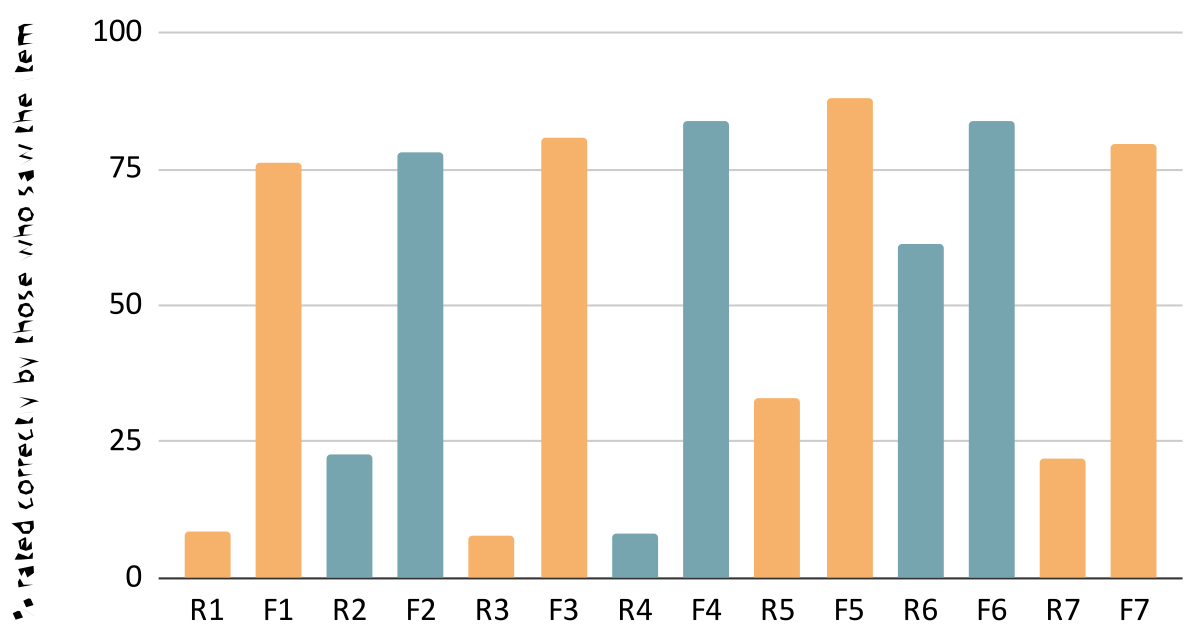

Figure 3: Comparison of what percentage of participants who saw Study 1 text messages with similar scenarios rated them correctly (Figure 1). A majority who saw the government-entity messages (R6, the "Amber Alert" message, and F6, the "tax audit and asset freeze" message) rated them correctly.

For Study 2, we hired Prolific to recruit a panel of 1,100 U.S. mobile phone users aged 18 or older. This enabled us to quickly collect data at a lower cost than Qualtrics, with 10% over the target 1,000 to allow for quality checks. To encourage completions, we shortened it to 6-8 minutes. Our IRB approved the revision. See Appendix D for the survey protocols.

### 3.3 Procedures

After the team reviewed and approved the Study 1 questionnaire, we provided Qualtrics with the URL to the coded online survey. Qualtrics distributed this link to its third-party panel providers. Participants who clicked on the survey URL and consented to participate were directed to a page to provide general demographic information and mobile phone usage. Those who met demographic quotas, were under 18, did not own a mobile phone, or failed CAPTCHA [19], were redirected to the Exit screen to ensure quality responses. The Study 1 questionnaire collected responses from June 26 to July 1, 2023. After meeting the quotas, a team member downloaded and visually inspected the responses for patterns and typed answers. Responses with repetitive copy-pastes were deleted. The master dataset was cleaned, prepped for analysis, and uploaded to a secure repository.

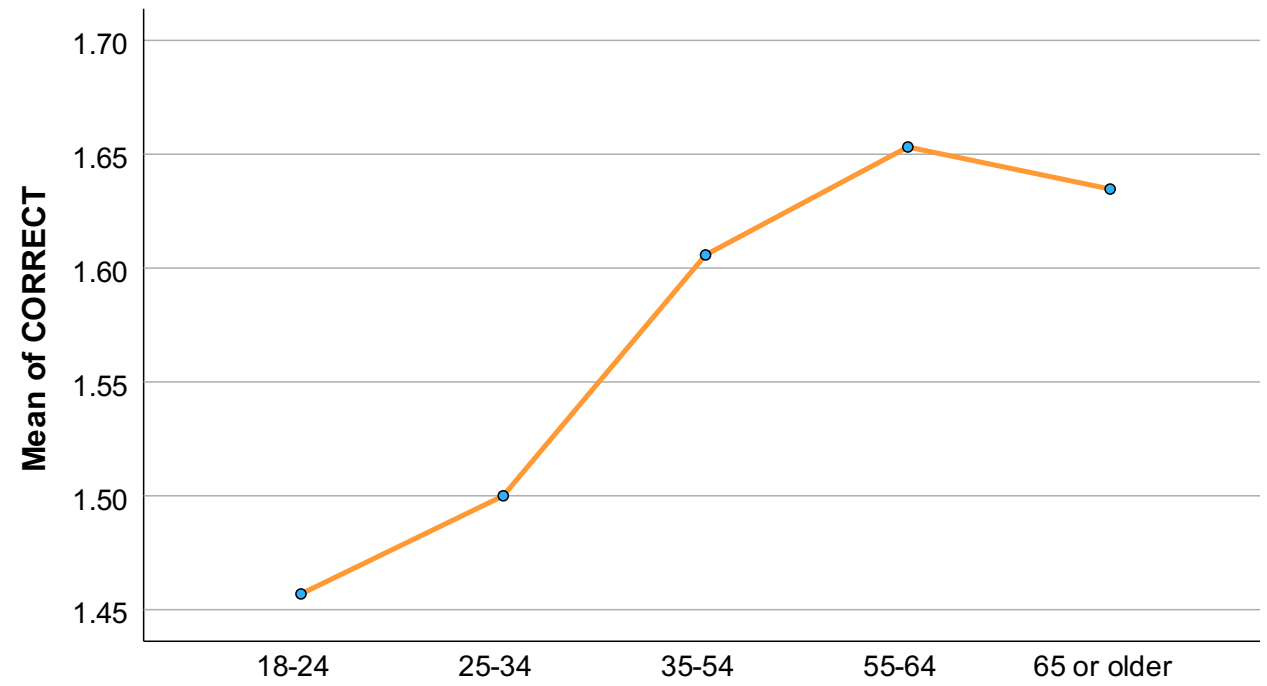

Figure 4: In Study 1, as age bracket increased up to 55-64, so did the average number of correct ratings of the simulated text messages (Adj. $R^2$=.006, $\beta$=0.084, $p$<.01).

Table 2: Regression analyses of demographic variables with respect to how well participants did at correctly identifying text messages. Results significant at the p<.050 level are bolded.

| *Variable predicting CORRECT* | *β* | *p-value* |
|---|---|---|
| GENDER_ALL | 0.067 | 0.188 |
| Female (vs. Not Female) | 0.066 | 0.219 |
| **AGE_ALL** | **0.084** | **0.007** |
| **18-24 (vs. 65 or older)** | **-0.635** | **0.043** |
| 25-34 (vs. 65 or older) | -0.073 | 0.858 |
| 35-54 (vs. 65 or older) | -0.076 | 0.768 |
| 55-64 (vs. 65 or older) | -0.086 | 0.776 |
| EDUCATION_ALL | -0.003 | 0.902 |
| Not in school, no 4-yr. deg. (vs. 4-yr. deg., no Phd) | -0.123 | 0.441 |
| **In school, no 4-yr. deg. (vs. 4-yr. deg., no Phd)** | **-0.653** | **0.003** |
| Phd (vs. 4-yr. degree, no Phd) | -0.309 | 0.189 |
| INCOME_ALL | 0.019 | 0.425 |
| $26,500 to $49,999 (vs. < poverty line) | 0.316 | 0.328 |
| $50K to $99,999 (vs. < poverty line) | 0.110 | 0.702 |
| $100,000 and up (vs. < poverty line) | 0.080 | 0.791 |
| Hisp/Lat/Spanish? | 0.024 | 0.455 |
| RACE/ETHNICITY_ALL | -0.044 | 0.166 |
| Black/African (vs. White/Caucasian) | -0.186 | 0.368 |
| Nat. Amer./Alaska Native (vs. White/Caucasian) | -0.484 | 0.473 |
| Asian - East or Central (vs. White/Caucasian) | -0.079 | 0.897 |
| Asian - South, SW., or SE. (vs. White/Caucasian) | 0.075 | 0.899 |
| Nat. Hawaiian/Pacif. Isl. (vs. White/Caucasian) | -21.464 | 1.000 |
| Mid. Eastern or N. African (vs. White/Caucasian) | -0.395 | 0.450 |
| Self-described (vs. White/Caucasian) | 0.550 | 0.364 |
| HOUSEHOLD_ALL | -0.033 | 0.294 |
| 1 ppl. (vs. 2 ppl.) | -0.316 | 0.098 |
| **3 ppl. (vs. 2 ppl.)** | **-0.436** | **0.026** |
| **4 ppl. (vs. 2 ppl.)** | **-0.511** | **0.004** |
| 5+ ppl. (vs. 2 ppl.) | -0.270 | 0.208 |

Table 3: Regression analyses of employment variables with respect to how well participants did at correctly identifying text messages. Results significant at the p<.050 level are bolded.

| *Variable predicting CORRECT* | *β* | *p-value* |
|---|---|---|
| WORKING_ALL | 0.060 | 0.058 |
| Stay-at-home parent (vs. Student-fulltime) | 0.308 | 0.453 |
| Part-time emp. outside home (vs. Student-fulltime) | 0.377 | 0.296 |
| Full-time emp. outside home (vs. Student-fulltime) | 0.517 | 0.108 |
| Unemployed (vs. Student-fulltime) | 0.315 | 0.366 |
| Self-employed (vs. Student-fulltime) | 0.719 | 0.104 |
| Student-parttime (vs. Student-fulltime) | 0.704 | 0.277 |
| Retired (vs. Student-fulltime) | 0.689 | 0.052 |
| Disabled (vs. Student-fulltime) | 0.457 | 0.355 |
| **JOB_ALL** | **0.143** | **0.006** |
| **Business/Fin. Ops** (vs. Constr. and Extraction) | **1.125** | **0.014** |
| Computer/Math. (vs. Constr. and Extraction) | 0.682 | 0.139 |
| **Office/Admin. Support** (vs. Constr. and Extr.) | **1.535** | **0.001** |
| Sales and Related (vs. Constr. and Extraction) | 0.655 | 0.143 |
| **Healthc. Pract., Tech.** (vs. Constr. and Extr.) | **1.671** | **<.001** |
| **Educ. Instruct., Library** (vs. Constr. and Extr.) | **1.507** | **0.002** |
| **Food Prep. and Service** (vs. Constr. and Extr.) | **1.344** | **0.006** |
| **Management** (vs. Constr. and Extr.) | **1.594** | **0.002** |

For Study 2, we advertised on Prolific for a "Study of Message Attitudes and Judgments," with a URL for access to our questionnaire. This simplified survey contained checks for age 18 or older, mobile phone use, and CAPTCHA, and ran on Feb. 2, 2024. We prepared the dataset similarly to above.

### 3.4 Participants

For Study 1, Qualtrics sourced responses from 1,000 people plus 1% overage. All had passed the attention-check item 2/3 of the way through the survey that directed them to answer with the 4th bullet point to retain their responses. The deletion of four responses with evidence of repeated nonsense copy-pastes into a text-input box resulted in a total dataset of *N*=1,007 (Table 1). A post-hoc analysis using G*Power found that this study has 95%-100% power to detect a small effect size (0.15) in linear regressions, consistent with the sample size and small effect sizes found in Sheng et al. [34].

About 75% reported receiving at least "a little" security training, and about one-third reported training specifically to help "identify fraudulent links or other threats in text messages." About half (51.5%) reported spotting and avoiding SMiSh in the past three months. Other responses were: "No, I have not noticed any fraudulent links in email, text messages, or web posts" (15.4%), "Yes, but it turned out to be a test being conducted as part of security awareness training" (7.0%), "Yes, and it turned out to be a scam, but nothing bad happened, to my knowledge" (15.7%), "Yes, and it turned out to be a scam, and I suffered a bad outcome (such as malware or theft of account credentials)" (4.4%), and Not Sure (6.1%).

For Study 2, data cleaning yielded *N*=1073 responses from Prolific (Table 6). A post-hoc analysis using G*Power found that this study has 100% power to detect a small effect size (0.15) in linear regressions and 95% power to detect the same effect size in chi-square tests, consistent with the sample size and small effect sizes found in Sheng et al. [34].

About 71% had received "a little" security training. One-quarter reported training to "identify fraudulent links or other threats in text messages." More than two-thirds (67.4%) reported spotting and avoiding SMiSh within the past three months. Other responses were: "No, I have not noticed any fraudulent links in email, text messages, or web posts" (19.3%), "Yes, but it turned out to be a test being conducted as part of security awareness training" (1.4%), "Yes, and it turned out to be a scam, but nothing bad happened, to my knowledge" (8.0%), "Yes, and it turned out to be a scam, and I suffered a bad outcome (such as malware or theft of account credentials)" (1.4%), and Not Sure (2.5%).

Table 4: Regression analyses of mobile phone variables with respect to how well Study 1 participants did at correctly identifying text messages. Results significant at the p<.050 level are bolded.

| *Variable predicting CORRECT* | *β* | *p-value* |
|---|---|---|
| PHONE_ALL | 0.045 | 0.157 |
| iOS smartphone (vs. Android smartphone) | 0.147 | 0.498 |
| Other type of smartphone (vs. Android smartphone) | -1.242 | 0.143 |
| Feature phone (vs. Android smartphone) | 0.145 | 0.891 |
| No-camera basic phone (vs. Android smartphone) | 19.045 | 0.999 |
| USAGE_ALL | 0.007 | 0.814 |
| <6 hours (vs. 11-20 hours) | -0.011 | 0.974 |
| 6-10 hours (vs. 11-20 hours) | 0.373 | 0.207 |
| 21-30 hours (vs. 11-20 hours) | 0.094 | 0.770 |
| **>30 hours (vs. 11-20 hours)** | **0.580** | **0.046** |

Table 5: Regression analyses of security-relevant variables with respect to how well participants did at correctly identifying text messages. Results significant at the p<.050 level are bolded.

| *Variable predicting CORRECT* | *β* | *p-value* |
|---|---|---|
| Self-reported amt. of experience working w/sensitive data | 0.001 | 0.971 |
| Self-reported amt. of training in spotting fraudulent messages | -0.022 | 0.481 |
| Self-reported amt. of overall security awareness training | -0.039 | 0.054 |
| **Security Behavior Intention** | **-0.162** | **<.001** |
| Security Attitude | -0.014 | 0.654 |
| **Freq. of security breaches - personal experiences** | **-0.065** | **0.024** |
| Freq. of security breaches - experiences of close ties | -0.004 | 0.889 |
| Freq. of security breaches - heard or read about them | 0.006 | 0.826 |
| Fell for a scam message w/in 3 mos., but it was a test (vs. recognized and avoided scam) | -0.270 | 0.290 |
| Fell for a scam message w/in 3 mos., but no harm (vs. recognized and avoided scam) | -0.021 | 0.909 |
| **Fell for a scam message w/in 3 mos. and suffered harm (vs. recognized and avoided scam)** | **-1.089** | **0.001** |
| Have not noticed any scam messages w/in 3 mos. (vs. recognized and avoided scam) | 0.025 | 0.894 |
| Not sure (vs. recognized and avoided scam) | -0.426 | 0.117 |

### 3.5 Data Analysis

We used IBM SPSS and Google Sheets to calculate descriptive and inferential statistics and to create figures. Regressions assessed the extent to which predictors significantly accounted for variances in rating correctness, using model fit and a 95% confidence interval to determine statistical significance. To score the SMiShing assessment (Section 3.1), we counted as correct any answer for a simulated "fake" text message (F1-7) rated "Fraudulent" or "Likely Fraudulent," and any answer for a simulated "real" text message (R1-7) rated "Legitimate" or "Likely Legitimate." This scheme computed an outcome variable, CORRECT, used in regressions. For CORRECT, we assigned values between 3 and 0 based on whether the participant rated three, two, one, or zero messages correctly.

In Study 2, we simplified the message rating to "Real - this looks like a legitimate message," or "A scam - this looks like a fraudulent message." This was used to compute the outcome variable CORRECT. We performed chi-squares and regressions to assess whether variables significantly accounted for variances in "real" and "fake" rating correctness.

## 4. Study 1 results

### 4.1 RQ1: Accuracy in IDing Simulated SMiSh vs. 'Real' Texts

Overall, participants correctly identified whether the messages were legitimate or fraudulent 52.6% of the time, calculated by dividing the number of correct ratings (Fraudulent/Likely Fraudulent for the "fake" messages, and Legitimate/Likely Legitimate for the "real" ones) by total number of messages seen. We found that participants did much better

Real 1: Real URL in body text, real short code is the sender

Real 2: Real phone number in body text, real short code is the sender

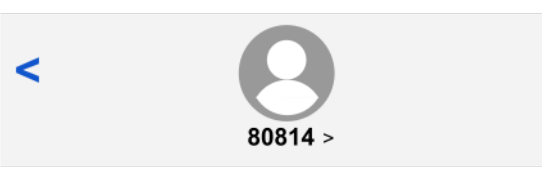

Real 3a: "#" prefix / reply text, real short code is the sender

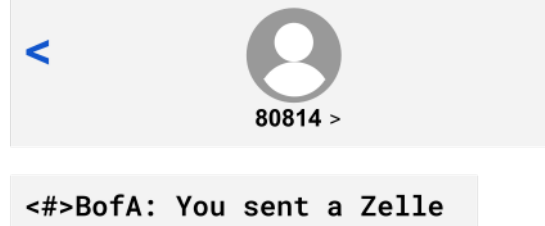

Real 3b: "#" prefix / call us text, real short code is the sender

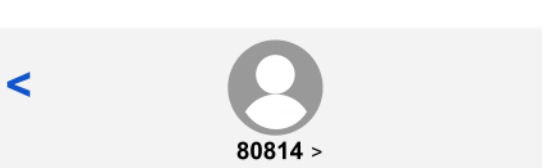

Fake 1a: Fake URL in body text, fake email is the sender

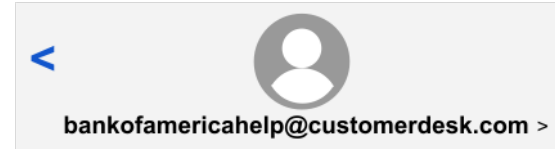

Fake 2a: Fake phone number in body text, fake email is the sender

Fake 3a: "#" prefix in body text, fake email is the sender

Fake 1b: Fake URL in body text, fake phone number is the sender

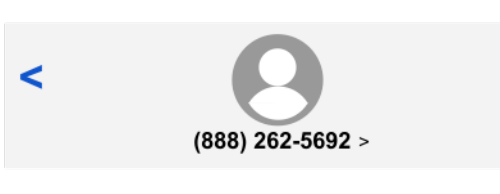

Fake 2b: Fake phone number in body text, fake number is the sender

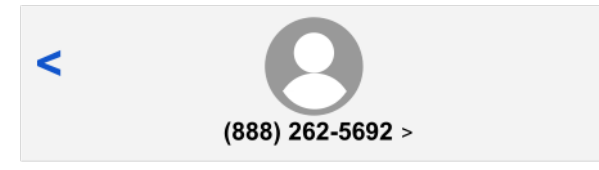

Fake 3b: "#" prefix in body text, fake phone number is the sender

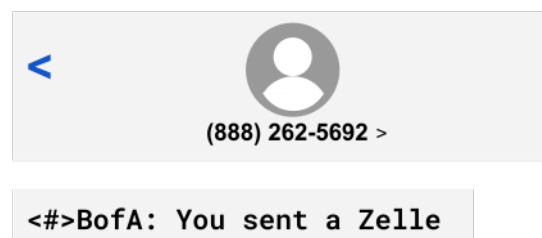
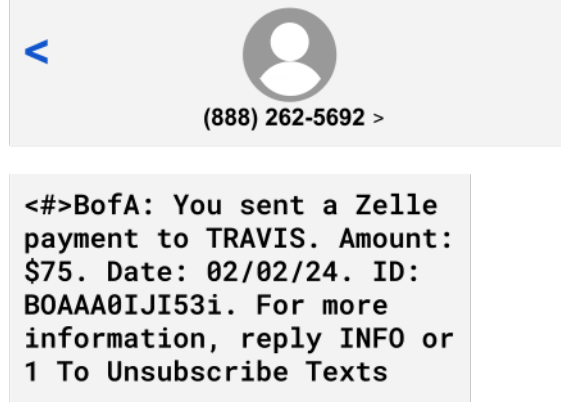

Fake 1c: URL in body text, fake short code is the sender

Fake 2c: Fake phone number in body text, fake short code is the sender

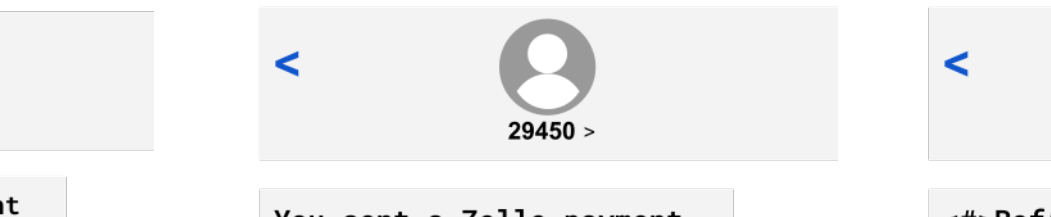

Fake 3c: "#" prefix in body text, fake short code is the sender

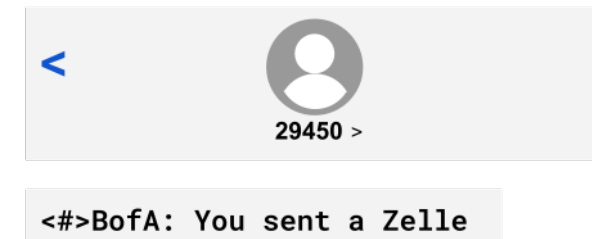

Figure 5:

In Study 2, we randomly served three of 13 variations of the same text message to survey participants. Each saw a randomly chosen one of four "real" or legitimate versions, using real URLs, phone numbers, body copy language, and short codes from Bank of America (top row).

Each also saw a randomly chosen two of nine "fake" or fraudulent versions, using non-BofA URLs, phone numbers, body copy language, and a short code that is not assigned to any real business.

at correctly identifying the simulated “fake” text messages (81.4%) than at correctly identifying the “real” ones (23.5%) (Figure 2). We did not find a significant overall effect by whether participants indicated that they had an account with the message entity. Most participants reported receiving at least “a little” security training, which may have contributed to high rates at which they could spot the SMiSh. However, it may have also led them to over-correct and misidentify legitimate messages, as happened in Sheng et al.’s study of phishing vulnerability and educational outcomes [34] and Timko et al.’s study of SMiShing susceptibility [37].

Among the simulated “real” text messages, only R6 (the simulated “Amber Alert” text message) was correctly identified as Legitimate /Likely Legitimate by a majority (61.3%) of those who saw it, followed by R5 (with the Phone Contact identifier and link to a popular video platform) (32.9%). Among the simulated “fake” messages, participants did the best at correctly rating F5 (with the cryptic suggestion that the receiver’s face was identifiable in nude images) as Fraudulent /Likely Fraudulent (88.0%). Participants tended to reply “Not Sure” more often for the simulated “real” text messages than for the simulated “fake” ones (Figure 3), suggesting that there were fewer interface indicators available to guide their judgments about legitimacy. However, the lack of consistent and directly comparable variations among the text messages left us unable to test for statistical significance in the message comparisons and completely answer RQ1. We deployed a follow-up survey to collect this data (Section 5).

When asked how they would respond to any given simulated “fake” message, a minority of participants indicated they would report the message using device options such as clicking Block This Caller or Report Junk (38.7%), while a majority said they would delete it and/or ignore it (73.3%). Responses were similar for the “real” messages (25.4% and 61.3%, respectively), which participants often incorrectly identified as SMiSh or likely SMiSh. While relatively few people selected “Reply to SMS text message to provide information” for the simulated “fake” messages (5.3%), some indicated that they would reply with STOP, BLOCK or other codes (17.5%). This still may accomplish the goal of the SMiSh attacker, since they may be testing the number to see if it remains in service and would be useful for a future scam [30]. Few participants said that they would respond in other ways that could meet the attacker’s goals: click on the link (6.3%), forward the message to someone else (3.3%), or keep, save or archive the message (4.9). A minority said they would check the link on device, either by copy-pasting or typing the link into their phone’s web browser (11.4%). Checking the link is recommended for phishing detection and mitigation [3, 12, 34], but is more easily done on a larger device.

In the “Other” box, some participants typed in that they would check with the message entity, such as their bank, to verify that the message was real and/or that the scenario reflected their recent account activity. See Appendix C for a complete list of “Other” responses from participants.

When asked why they would respond a certain way, participants’ responses were similar for the fraudulent messages as for the legitimate ones on four measures: sense of urgency (13.0% for "fake" vs. 14.2% for "real"), curiosity (10.9% for "fake" vs. 12.6% for "real"), seeking a good outcome for myself (12.3% for "fake" vs. 12.7% for "real"), and lack of interest in the message (36.4% for "fake" vs. 32.1% for "real"). Even for the legitimate messages, participants reported little trust in the sender (14.1%, versus 10.3% for the "fake" messages) or in the link URL (9.6%, versus 7.7% for the "fakes"), suggesting that these source indicators were only of marginal help in participants' assessments. Slightly more than half of participants who saw fraudulent messages reported "seeking to avoid a bad outcome for myself" as reasons for their response (50.7%), though a significant minority also reported this for the legitimate messages (47.0%).

Table 6: For Study 2 participants, counts for demographics, mobile phone and usage, experience working with sensitive data, and jobs.

| Age | | Education | | Household Inc. | | Gender Identity | | Hisp./Lat./Sp.? | | Racial/Ethnic Identity | | Household Size | |
|---|---|---|---|---|---|---|---|---|---|---|---|---|---|
| 18-24 | 137 | No 4y deg. and not in school | 429 | < $26.5K poverty line | 192 | Female | 526 | No | 1009 | White/Cauc. | 838 | 1 ppl. | 236 |
| 25-34 | 194 | No 4y deg., but in school | 52 | $26.5-$49K | 221 | Male | 524 | Yes | 60 | Black/African | 141 | 2 ppl. | 383 |
| 35-54 | 353 | 4y deg., but no doctorate | 550 | $50-$99K | 385 | Nonbinary | 18 | Prefer not to say | 4 | Asian – total for all regions | 66 | 3 ppl | 195 |
| 55-64 | 227 | Has doctorate | 42 | $100K+ | 275 | Self-described | 3 | | | Native Am. or Alaska Native | 2 | 4 ppl. | 156 |
| 65+ | 162 | | | | | Prefer not to say | 2 | | | Self-described | 17 | 5+ ppl. | 102 |
| | | | | | | | | | | Prefer not to say | 9 | | |

| Primary Mobile Phone | | Usage / Past Week | | Exp. w/Sens. Data | | Primary Job Status | | Top Occupations (FT or PT) | |
|---|---|---|---|---|---|---|---|---|---|
| Smartphone - Android | 565 | <6 hrs. | 157 | None at all | 425 | Full-time (FT) | 519 | Computer and Mathematical | 83 |
| Smartphone - iOS/Apple | 497 | 6-10 hrs. | 250 | Only a little | 266 | Part-time (PT) | 180 | Sales and Related | 76 |
| Other smartphone | 3 | 11-20 hrs. | 266 | A moderate amount | 179 | Unemployed | 90 | Educational Instr. and Library | 67 |
| Featurephone with camera | 6 | 21-30 hrs. | 194 | A lot | 100 | Retired | 142 | Business and Fin. Operations | 65 |
| Basic phone with no camera | 2 | >30 hrs. | 206 | A great deal | 103 | At-home parent | 42 | Arts, Desg., Ent., Sports, Media | 53 |
| | | | | | | Other | 61 | Management | 49 |
| | | | | | | FT student | 36 | Office and Admin. Support | 46 |
| | | | | | | PT student | 3 | Healthcare Support | 38 |
| | | | | | | | | Healthcare Pract. and Technical | 32 |

### 4.2 RQ2 and RQ3: Differences Among Comparison Groups

We found that variances in CORRECT could be significantly explained by participants' age (Figure 4), by whether they reported currently studying for a four-year degree (Table 2), by household size of 3-4 people (Table 2), and by use of a mobile phone for more than 30 hours per week (Tables 4-5). We also found that variances in CORRECT could be significantly explained by participants' job category (Table 3), with those in Construction and Extraction performing the worst and those in Educational Instruction and Library performing the best. We compare these with prior findings in Section 6.2.

We found that variances could be significantly accounted for by the frequency of their personal experiences of security breaches; their score on the Security Behavior Intentions Scale (SeBIS) [16] subscale for Proactive Awareness; and whether they had fallen for a scam message in the past three months and suffered a bad outcome (Table 5). We found no significant associations at the $p$<.05 level for the following variables: frequency of close ties experiencing security breaches, frequency of hearing or seeing news about security breaches, amount of experience working with sensitive data, whether they reported clicking on SMiSh in the past three months (but it was either a test or they suffered no apparent harm), whether they had not noticed any scam messages in the past three months, whether they had received either overall security awareness training or specific training on spotting and dealing with fraudulent text messages, and score on the SA-6 Security Attitude scale [20].

## 5. Study 2 Results

### 5.1 Comparisons of Variants on a Single Message

To zero in on message elements that mattered, we designed 13 variants on a Zelle notification from Bank of America -- four that use content cues to legitimate company messages and nine that use cues to fraudulent texts (Figure 5). We kept these very similar to real-life BofA messages (such as using the real short codes, URLs, phone numbers, and "#" prefix commonly seen in such texts) and to real-life Zelle scam messages (such as the phony email, phone numbers and reply text, and the non-BofA short code). Participants could have checked some content cues in the messages against materials on the BofA website, such as its official short codes in use, although the bank does not share examples of legitimate SMS messages that it sends as part of operations.

On average, Study 2 participants correctly identified whether the messages were legitimate or fraudulent 63.8% of the time.

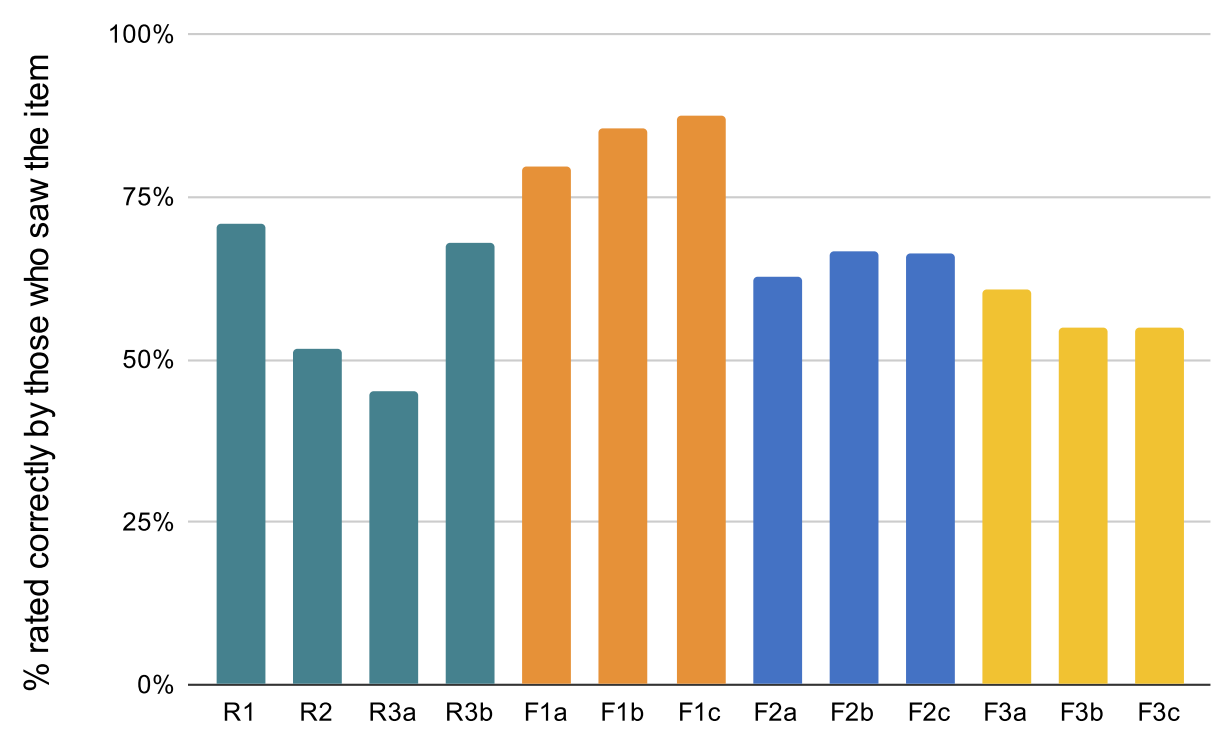

Figure 6: Comparison of what percentage of participants who saw Study 2 message variants rated them correctly. A majority who saw the nine "fake" variants rated them correctly (F1a-3c), while results for the four "real" variants (R1-3b) were mixed.

Table 7: Chi-square tests showed which of the pairwise comparisons between correct "fake" and "real" messages were significant, for participants who saw both messages.

| *"Fake"* | *Comparison with "Real" messages* |
|---|---|
| **All fake** | **All real: $x^2$=77.699, $p$<.001** |
| Fake 1a | (N.S. for all) |
| Fake 1b | (N.S. for all) |
| Fake 1c | (N.S. for all) |
| **Fake 2a** | R1: N.S., **R2: p<.001**, **R3a: p<.01**, R3b: N.S. |
| **Fake 2b** | **R1: p=.01**, R2: N.S., R3a: N.S., R3b: N.S. |
| **Fake 2c** | **R1: p<.005, R2: p<.005**, R3a: N.S., R3b: N.S. |
| **Fake 3a** | R1: N.S., R2: N.S., **R3a: p<.001, R3b: p<.05** |
| **Fake 3b** | R1: N.S., **R2: p<.05, R3a: p<.001, R3b: p<.001** |
| **Fake 3c** | R1: N.S., **R2: p<.05, R3a: p<.001, R3b: p<.05** |

Table 8: Among the scam Zelle messages shown in Study 2, many in the Age 18-24 and In College demographics fell for the messages if they displayed a fake phone number (Fake 2a-2c) or a fake short code (Fake 3a-3c) as the sender.

| | Age 18-24 | | 4-Year College Student | |
|---|---|---|---|---|
| | *Looks real* | *Looks fake* | *Looks real* | *Looks fake* |
| Fake 1a | 15.40% | 84.60% | 25.00% | 75.00% |
| Fake 1b | 9.09% | 90.90% | - | 100.00% |
| Fake 1c | 9.38% | 90.62% | 11.76% | 88.24% |
| Fake 2a | **50.00%** | **50.00%** | **70.00%** | **30.00%** |
| Fake 2b | 40.63% | 59.38% | **50.00%** | **50.00%** |
| Fake 2c | **59.26%** | **40.74%** | 45.45% | 54.54% |
| Fake 3a | **51.85%** | **48.15%** | **54.54%** | **45.45%** |
| Fake 3b | **55.56%** | **44.44%** | **60.00%** | **40.00%** |
| Fake 3c | **52.94%** | **47.06%** | **50.00%** | **50.00%** |

Table 9: Comparisons among susceptibility insights for phishing and SMiShing from prior work with the results of our studies

| Demographic | Susceptibility Found in Prior Work | Susceptibility Found in Our Studies |
|---|---|---|
| **Age** | Younger adults (18-25) [35, 40] and older adults [28] for phishing, 45-54 age group for smishing [31] | Younger adults (18-24) struggled to identify scam messages and sometimes misidentified real messages as fake. |
| **Gender** | Women, Men (more likely to distinguish phishing websites) [25, 28, 35] | No significant differences by gender. |
| **Education** | Higher education correlates with higher smishing success but doctorates struggled [31], Knowledge reduces phishing risk and link clicking [35, 40] | Those with no four-year degree but in school identified legitimate messages as fake, struggled to identify scam messages correctly. |
| **Training and Prior Experiences** | Prior exposure to anti-phishing education is generally associated with less susceptibility [31]. However, training may improve ability to avoid clicking on links, but not entering information on phishing sites [35, 40] | No significant effects found for self-reported amounts of training in spotting fraudulent messages or for security awareness training. Some significant negative associations between correctness and personal experiences, either of actual security breaches or news about security breaches. |
| **Employment** | Employment department and position [28] | Study 1 participants working in healthcare, education, office-related, and food-service job categories did significantly well in identifying real and scam messages. No significant effects found for other job categories. |

As in Study 1, however, they did significantly better at correctly identifying the simulated SMiSh (68.72%) than the simulated real messages (58.8%) (Figure 6, Table 7, overall $p<.001$). We found no significant effect for having a Bank of America account, though we did find that using Zelle was significantly associated with correctly identifying the scam messages (Adj. $R^2$=0.003, $\beta$=.092, $p<.05$). Among the simulated "real" text messages, two performed moderately well (R1, the "real url" body text, 70.8%, and R3b, the "# prefix / call us" body text, 67.9%), and two less so (R2, the "real phone number" body text, 51.5%, and R3a, the "# prefix / reply INFO" body text, 45.15%). The latter may have underperformed because they were more like SMiSh that use fake phone numbers and reply text.

Among the simulated "fake" messages, participants did better at correctly rating F1a-c (fake URL in body text) as fake (average 84.1%), and worse at F3a-c (enticement to reply to unsubscribe, average 56.9%) and F2a-c (fake phone number, average 65.2%). The "Reply to Unsubscribe" verbiage has shown up in many SMiSh, including a Zelle scam described on Brian Krebs' security blog [40] in which the SMS is followed by a faked security team call. The use of the "#" prefix, common to observed BofA official text messages [36], also may have been a primary reason for participants' incorrect ratings of F3a-c as "real".

#### 5.1 Effects of Age, Education, Experiences, and Training

Next, we examined the performance of the Age 18-24 and the 4-Year College Student demographic groups. A majority of these group members correctly rated all four "Real" messages. However, we found that being Age 18-24 was negatively associated with correctly identifying "Real" messages when controlling for both being a Bank of America customer and a Zelle user (Adj. $R^2$=0.016, $\beta$=-0.130, $p<.05$). The same negative effect was found for 4-Year College Students when controlling for these variables (Adj. $R^2$=0.003, $\beta$=-0.095, $p<.05$). It may be that past scams have made these participants so risk-averse that their default is to think that any Zelle message involving Bank of America is fake.

Both groups had mixed success in identifying the nine "Fake" messages (Table 8). Only the messages sent from a fake email address (Fake1a-1c) were consistently identified as a scam by at least half of those who saw it. However, the differences in correctly IDing the fake messages were non-significant.

Finally, we examined the effect of prior experiences and training. We found that, when controlling for both being a Bank of America customer and a Zelle user, having fallen for a scam message within the past three months (with any outcome) was associated with correctly rating all seen messages (Adj. $R^2$=0.008, $\beta$=0.097, $p$=.001), particularly the "Fake" messages (Adj. $R^2$=0.014, $\beta$=0.114, $p<.001$). Further, with the same controls, experience working with sensitive data was negatively associated with correctly rating the "Real" messages (Adj. $R^2$=0.003, $\beta$=-0.063, $p<.05$). When controlling for the bank and Zelle usage, hearing or seeing a lot about online breaches was negatively associated with correctly rating "Fake" messages (Adj. $R^2$=0.006, $\beta$=-0.065, $p<.05$).

We found no significant effects for the self-reported amount of training in spotting fraudulent messages or for overall security awareness training.

## 6. Discussion

This paper provides critical insights from two large-scale, representative samples into the demographic and other individual factors associated with susceptibility to smishing in U.S. adult mobile users. The header information, which is often used to help email users to identify phishing, did not help Study 1 and Study 2 participants to reliably and correctly identify "Fake" messages in a mobile context. This is consistent with results from Tabassum et al. [36] that participants looked at message content rather than sender information to judge possible SMiSh. Similarly to Timko et al. [37], our participants did worse than expected at correctly rating the simulated “Real” mobile messages. This should worry any entity (such as a government agency or a company) that needs to use mobile messaging to communicate urgent and important information.

### 6.1 Putting Our Results into Context

Our findings indicate that younger adults (18-24) are particularly vulnerable, struggling both to identify legitimate messages as genuine and to recognize fraudulent messages as scams. This contrasts with previous studies in phishing that suggest older adults are more susceptible [27] but aligns with other studies showing that younger adults are more susceptible to phishing [34, 39]. For SMiShing, the Rahman et al. study found the 45-54 age group were more susceptible to fraudulent messages, while the youngest participants were the least susceptible [30]. This indicates that age-related vulnerabilities are complex and nuanced for both phishing and SMiShing. The differences among research studies may be due to the specific messages used in simulations and whether participants are responding in a controlled study context.

Contrary to some prior research indicating gender differences in susceptibility to phishing [24, 27, 34], our study found no significant differences for the simulated SMiShing tasks. It may be that underlying factors, such as efforts to broaden participation in computing and to boost cybersecurity awareness and education, have equalized the gender differences found in previous studies.

Beyond demographics, our work identified other significant factors. In Study 1, participants working in domains such as Healthcare and Education were better at judging simulated text messages. This suggests that expertise in carefully perceiving and judging information credibility is protective against smishing. Previously, Ngo et al. [27] had found employment department and position to be protective. In Study 2, hearing or seeing a lot about online security breaches was negatively associated with correctly rating the "Fake" messages about Zelle transactions from Bank of America. This may be due to several factors including increased risk perception, fatigue or overconfidence.

In both studies, meanwhile, we found significant negative associations between correctly judging messages and participants’ education status – specifically, for those currently in school for a four-year degree. Rahman et al. also found doctorate-holders to be susceptible to SMiShing [30]. Our anecdotal conversations with industry researchers are that they are seeing the same patterns, but that, like us, they do not have a good explanation for the results. One explanation is that academic work requiring sustained, focused attention results in cognitive fatigue, leaving little energy or focus remaining for secondary tasks such as message judgments. The undergraduate students in our lives have suggested that the sheer volume of unsolicited messages that they receive (whether spam or malicious) wears down their ability to detect what is real and what is fabricated.

### 6.2 The Need for New Ideas for Helping Mobile Users

Perhaps most importantly, our study found no significant effects for self-reported training in spotting fraudulent messages or for overall security awareness training. This is particularly concerning, as it suggests that current security messaging and training is falling short of helping people strike the right balance in their judgments of smishing. Our findings contrast with the general assumption that training is beneficial for SMiShing mitigation [30] and previous work suggesting that anti-phishing education is associated with less susceptibility [34, 39].

Based on our findings, we recommend that industry and academic experts plan more-targeted risk-awareness campaigns and educational interventions. We think a focus on young adults and college students will pay off, as our results show they are particularly vulnerable. What these groups learn early about avoiding mobile scams will help them form good security habits for years to come.

Our data suggests that new ideas are needed, as we found no significant effect of self-reported amounts of either security awareness training or specific training in spotting fraudulent messages. In Study 2, however, we did find evidence that falling for a scam message within the previous three months was protective, regardless of whether it was a test or whether it caused harm.

One solution, then, could be adapting trainings for email such as Phish Me [8] to a mobile context. These tools send simulated phishing attack emails as a real-life test of an employee’s susceptibility to scam messages, often showing an educational message when someone clicks on the URL. A similar intervention for mobile, using the type of content such as Zelle messages that a particular demographic group is likely to encounter, could anchor the experience similarly in the memory of mobile users and help them identify future scam messages. We recommend that such a tool for mobile message training also incorporate simulated real messages, to help train users also in how to spot a legitimate communication and avoid defaulting to judging everything as fake.

As a training backstop and additional layer of mitigation, we also recommend that U.S. regulators and telecom providers work with usability experts to improve verification and interface aids for mobile messaging. Adding icons for verified senders and warning labels for external senders, like those now used in email systems, would help mobile users to more easily identify trusted sources. It could reduce the frequency of scammers tricking mobile phone users by faking the name of a well-known entity with a subtle misspelling, such as "Amazom" or "Facebo0k Security."

Finally, we suggest more research to explore the relationships among customers' age, educational attainment, job experience, security training, and SMiShing vulnerability for financial providers who are often impersonated by scammers. Our results and those of prior work show that there is no simple answer to how to allocate scarce resources for security-awareness training and mitigation tools. New approaches may be needed based on not only age of customers but whether they are in college or hold a doctorate, whether they work in a job that helps them practice their skills for judging information, and other individual factors.

## 7. Limitations and Future Work

Our two studies provide useful statistical data for assessing how well U.S. participants can distinguish fraudulent from legitimate text messages. However, this cross-sectional design is insufficient to establish cause-and-effect relationships. Our survey contract terms did not allow us to collect contact information for follow-up. We recommend conducting in-depth interview studies to gain more context on how and why people rate simulated text messages as fraudulent or legitimate, and to understand the recency and extent of their security-relevant training.

A future follow-up study might also create pairs of real and fake text messages that are identical except for one attribute, to more precisely determine which attribute is the most likely to lead people to fall for SMiSh. Such a study should also examine details of account knowledge's influence on participants' ratings of messages as either legitimate or fraudulent, such as knowledge that you have received a similar message in the past from the message entity or that you have reason to believe that the message content is referencing a real-life transaction on your account.

Both studies' results suggest that users need help to identify legitimate messages more readily. In future work, we will test interface design improvements for mobile and wearable interfaces, such as indicators or a naming scheme for the SMS short codes, that can provide prominent cues to which SMS text messages are from legitimate sources.

We practiced a careful method of iterative survey development to ensure our questionnaires' clarity and comprehensibility, and the anonymous method encouraged full honesty in answers. However, like all survey studies, ours is subject to several biases, such as self-report bias and social desirability bias, that possibly have skewed the results. A replication of the Study 1 survey in a similarly U.S. representative sample would help to validate the results and interpretations of this data. Such a replication would also benefit from separating out a closed-ended item for rating of whether the messages are legitimate or fraudulent from one that measures the participant's confidence in their rating. This would help validate our method of translating the interval confidence rating into a binary correctness variable and provide more clarity about participants' cognitive assessments.

Finally, we developed a useful way to simulate SMiSh without sending participants unsolicited text messages that could have panicked them or led them to feel tricked once debriefed. As is common in simulation studies, participants lacked direct consequences for their ratings and responses, such as missing opportunities from being too risk-averse in their clicks [34], which may have biased results. This could explain some of the differences between our findings and prior work. In a future study, we may explore how to conduct a more true-to-life SMiSh deception test similar to Rahman et al. [30] that boosts ecological validity. For such work, we will seek advice from our Information Security Office and/or from our Institutional Research Board on best practices, such as making the ISO aware of our study so that they can answer questions and provide reassurance to anyone who contacts them worried about the content of our messages. We will also pilot the study to make sure that the messages are not so alarming as to outweigh the benefit gained by the research.

## 8. Conclusions

In these studies, we collected and analyzed data from two large-scale panels of U.S. adult mobile phone users. We found that younger people and college students were signifi-cantly more vulnerable to SMiSh, that participants overall struggled to identify legitimate text messages, and that some participants were misled if the fraudulent text messages men-tioned an entity that they thought they had an account with. We also found that a previously documented gender differ-ence in vulnerability has disappeared. Our study contributes novel and badly needed knowledge of demographic suscepti-bility to scam messages for the era of mobile phones and widespread use of remote messaging, and examples of simu-lated "real" and "fake text messages and a survey protocol for use in research on SMiShing. Finally, we provide recommen-dations based in our data for use by researchers, regulators, and telecom providers. We hope these findings, and any fu-ture work based on it, will meaningfully improve the user ex-perience and security of the mobile internet.

## Acknowledgments

We are grateful to our industry collaborators for their assistance. This work was funded by the Center for Cybersecurity Analytics and Automation (established with NSF award #1822150). The first author was partially supported by NSF Grant No. 2346281, DOD Department of the Army (DA) Grant No. W911NF2410189, and a 2024 Google Research Award.

## References

[1] Hossein Abroshan, Jan Devos, Geert Poels, and Eric Laermans. 2021. COVID-19 and Phishing: Effects of Human Emotions, Behavior, and Demographics on the Success of Phishing Attempts During the Pandemic. *IEEE Access* 9, (2021), 121916–121929. https://doi.org/10.1109/ACCESS.2021.3109091

[2] Ali F. Al-Qahtani and Stefano Cresci. 2022. The COVID-19 scamdemic: A survey of phishing attacks and their countermeasures during COVID-19. *Iet Inf. Secur.* 16, 5 (September 2022), 324–345. https://doi.org/10.1049/ise2.12073

[3] Mohamed Alsharnouby, Furkan Alaca, and Sonia Chiasson. 2015. Why phishing still works: User strategies for combating phishing attacks. *Int. J. Hum.-Comput. Stud.* 82, (October 2015), 69–82. https://doi.org/10.1016/j.ijhcs.2015.05.005

[4] bferrite. 2017. The SMISHING threat – unraveling the details of an attack. *Check Point Blog*. Retrieved September 5, 2023 from https://blog.checkpoint.com/research/smishing-threat-unraveling-details-attack/

[5] Mark Blythe, Helen Petrie, and John A. Clark. 2011. F for fake: four studies on how we fall for phish. In *Proceedings of the SIGCHI Conference on Human Factors in Computing Systems* (*CHI '11*), May 07, 2011. Association for Computing Machinery, New York, NY, USA, 3469–3478. https://doi.org/10.1145/1978942.1979459

[6] Cristian Bravo-Lillo, Saranga Komanduri, Lorrie Faith Cranor, Robert W. Reeder, Manya Sleeper, Julie Downs, and Stuart Schechter. 2013. Your attention please: designing security-decision UIs to make genuine risks harder to ignore. In *Proceedings of the Ninth Symposium on Usable Privacy and Security* (*SOUPS '13*), July 24, 2013. Association for Computing Machinery, New York, NY, USA, 1–12. https://doi.org/10.1145/2501604.2501610

[7] Misty Casul. 2022. SMS Marketing Statistics You Should Know in 20222. *Thrive Internet Marketing Agency*. Retrieved September 5, 2023 from https://thriveagency.com/news/sms-marketing-statistics-you-should-know-in-2022/

[8] Cofense. PhishMe Security Awareness Training (SAT) Platform. *Cofense*. Retrieved February 13, 2025 from https://cofense.com/phishme-security-awareness-training-(sat)-platform

[9] Stacy Cowley and Lananh Nguyen. 2022. Fraud Is Flourishing on Zelle. The Banks Say It's Not Their Problem. *The New York Times*. Retrieved January 15, 2024 from https://www.nytimes.com/2022/03/06/business/payments-fraud-zelle-banks.html

[10] Shelby R. Curtis, Prashanth Rajivan, Daniel N. Jones, and Cleotilde Gonzalez. 2018. Phishing attempts among the dark triad: Patterns of attack and vulnerability. *Comput. Hum. Behav.* 87, (October 2018), 174–182. https://doi.org/10.1016/j.chb.2018.05.037

[11] Rachna Dhamija, J. D. Tygar, and Marti Hearst. 2006. Why Phishing Works. In *Proceedings of the SIGCHI Conference on Human Factors in Computing Systems* (*CHI '06*), 2006. ACM, New York, NY, USA, 581–590. https://doi.org/10.1145/1124772.1124861

[12] Julie S. Downs, Mandy B. Holbrook, and Lorrie Faith Cranor. 2006. Decision Strategies and Susceptibility to Phishing. In *Proceedings of the Second Symposium on Usable Privacy and Security* (*SOUPS '06*), 2006. ACM, New York, NY, USA, 79–90. https://doi.org/10.1145/1143120.1143131

[13] Julie S. Downs, Mandy Holbrook, and Lorrie Faith Cranor. 2007. Behavioral response to phishing risk. In *Proceedings of the anti-phishing working groups 2nd annual eCrime researchers summit*, October 04, 2007. ACM, Pittsburgh Pennsylvania USA, 37–44. https://doi.org/10.1145/1299015.1299019

[14] Serge Egelman, Lorrie Faith Cranor, and Jason Hong. 2008. You'Ve Been Warned: An Empirical Study of the Effectiveness of Web Browser Phishing Warnings. In *Proceedings of the SIGCHI Conference on Human Factors in Computing Systems* (*CHI '08*), 2008. ACM, New York, NY, USA, 1065–1074. https://doi.org/10.1145/1357054.1357219

[15] Serge Egelman, Marian Harbach, and Eyal Peer. 2016. Behavior Ever Follows Intention?: A Validation of the Security Behavior Intentions Scale (SeBIS). In *Proceedings of the 2016 CHI Conference on Human Factors in Computing Systems* (*CHI '16*), 2016. ACM, New York, NY, USA, 5257–5261. https://doi.org/10.1145/2858036.2858265

[16] Serge Egelman and Eyal Peer. 2015. Scaling the Security Wall: Developing a Security Behavior Intentions Scale (SeBIS). In *Proceedings of the 33rd Annual ACM Conference on Human Factors in Computing Systems* (*CHI '15*), 2015. ACM, New York, NY, USA, 2873–2882. https://doi.org/10.1145/2702123.2702249

[17] Serge Egelman and Eyal Peer. 2015. Predicting privacy and security attitudes. *ACM SIGCAS Comput. Soc.* 45, 1 (2015), 22–28.

[18] Emily Cahill. 2022. Phishing, Smishing and Vishing: What's the Difference? - Experian. Retrieved September 5, 2023 from https://www.experian.com/blogs/ask-experian/phishing-smishing-vishing/

[19] Cori Faklaris. 2021. Qualtrics RelevantID, reCAPTCHA, and other tips for survey research in 2021. *Cori Faklaris' blog - HeyCori*. Retrieved July 21, 2021 from https://blog.corifaklaris.com/2021/04/25/qualtrics-relevantid-recaptcha-and-other-tips-for-survey-research-in-2021/

[20] Cori Faklaris, Laura Dabbish, and Jason I Hong. 2019. A Self-Report Measure of End-User Security Attitudes (SA-6). In *Proceedings of the Fifteenth Symposium on Usable Privacy and Security (SOUPS 2019)*, August 12, 2019. USENIX Association Berkeley, CA, Santa Clara, CA, 18. Retrieved from https://www.usenix.org/system/files/soups2019-faklaris.pdf

[21] B.J. Fogg, Cathy Soohoo, David R. Danielson, Leslie Marable, Julianne Stanford, and Ellen R. Tauber. 2003. How Do Users Evaluate the Credibility of Web Sites?: A Study with over 2,500 Participants. In *Proceedings of the 2003 Conference on Designing for User Experiences* (*DUX '03*), 2003. ACM, New York, NY, USA, 1–15. https://doi.org/10.1145/997078.997097

[22] Jason Hong. 2012. The state of phishing attacks. *Commun. ACM* 55, 1 (January 2012), 74–81. https://doi.org/10.1145/2063176.2063197

[23] Mohammad S. Jalali, Maike Bruckes, Daniel Westmattelmann, and Gerhard Schewe. 2020. Why Employees (Still) Click on Phishing Links: Investigation in Hospitals. *J. Med.*

*Internet Res.* 22, 1 (2020), e16775. https://doi.org/10.2196/16775

[24] Tian Lin, Daniel E. Capecci, Donovan M. Ellis, Harold A. Rocha, Sandeep Dommaraju, Daniela S. Oliveira, and Natalie C. Ebner. 2019. Susceptibility to Spear-Phishing Emails: Effects of Internet User Demographics and Email Content. *ACM Trans Comput-Hum Interact* 26, 5 (July 2019), 32:1-32:28. https://doi.org/10.1145/3336141

[25] Sandhya Mishra and Devpriya Soni. 2019. SMS Phishing and Mitigation Approaches. In *2019 Twelfth International Conference on Contemporary Computing (IC3)*, August 2019. 1–5. https://doi.org/10.1109/IC3.2019.8844920

[26] Mattia Mossano, Oksana Kulyk, Benjamin Maximillian Berens, Elena Marie Häußler, and Melanie Volkamer. 2023. Influence of URL Formatting on Users' Phishing URL Detection. In *Proceedings of the 2023 European Symposium on Usable Security* (*EuroUSEC '23*), October 16, 2023. Association for Computing Machinery, New York, NY, USA, 318–333. https://doi.org/10.1145/3617072.3617111

[27] Fawn T. Ngo, Anurag Agarwal, and Katherine Holman. Cyber Hygiene and Cyber Victimization Among Limited English Proficiency (LEP) Internet Users: A Mixed-Method Study. *Vict. Offenders* 0, 0 , 1–22. https://doi.org/10.1080/15564886.2024.2329765

[28] Peter G. Polson, Clayton Lewis, John Rieman, and Cathleen Wharton. 1992. Cognitive walkthroughs: a method for theory-based evaluation of user interfaces. *Int. J. Man-Mach. Stud.* 36, 5 (May 1992), 741–773. https://doi.org/10.1016/0020-7373(92)90039-N

[29] Qualtrics. Survey Methodology & Compliance Best Practices. Retrieved December 10, 2023 from https://www.qualtrics.com/support/survey-platform/survey-module/survey-checker/survey-methodology-compliance-best-practices/

[30] Md Lutfor Rahman, Daniel Timko, Hamid Wali, and Ajaya Neupane. 2023. Users Really Do Respond To Smishing. In *Proceedings of the Thirteenth ACM Conference on Data and Application Security and Privacy* (*CODASPY '23*), April 24, 2023. Association for Computing Machinery, New York, NY, USA, 49–60. https://doi.org/10.1145/3577923.3583640

[31] E. M. Redmiles, A. R. Malone, and M. L. Mazurek. 2016. I Think They're Trying to Tell Me Something: Advice Sources and Selection for Digital Security. In *2016 IEEE Symposium on Security and Privacy (SP)*, May 2016. 272–288. https://doi.org/10.1109/SP.2016.24

[32] Elissa M Redmiles, Yasemin Acar, Sascha Fahl, and Michelle L Mazurek. 2017. *A Summary of Survey Methodology Best Practices for Security and Privacy Researchers*. University of Maryland, Technical Reports of the Computer Science Department. Retrieved from https://drum.lib.umd.edu/items/683d78b0-a0e3-4fae-9c93-b75aae4ad11b

[33] Elissa M. Redmiles, Sean Kross, and Michelle L. Mazurek. 2016. How I Learned to Be Secure: A Census-Representative Survey of Security Advice Sources and Behavior. In *Proceedings of the 2016 ACM SIGSAC Conference on Computer and Communications Security* (*CCS '16*), 2016. ACM, New York, NY, USA, 666–677. https://doi.org/10.1145/2976749.2978307

[34] Steve Sheng, Mandy Holbrook, Ponnurangam Kumaraguru, Lorrie Faith Cranor, and Julie Downs. 2010. Who falls for phish? a demographic analysis of phishing susceptibility and effectiveness of interventions. In *Proceedings of the SIGCHI Conference on Human Factors in Computing Systems* (*CHI '10*), April 10, 2010. Association for Computing Machinery, New York, NY, USA, 373–382. https://doi.org/10.1145/1753326.1753383

[35] Yunpeng Song, Cori Faklaris, Zhongmin Cai, Jason I. Hong, and Laura Dabbish. 2019. Normal and Easy: Account Sharing Practices in the Workplace. *Proc ACM Hum-Comput Interact* 3, CSCW (November 2019), 83:1-83:25. https://doi.org/10.1145/3359185

[36] Sarah Tabassum, Cori Faklaris, and Heather Richter Lipford. 2024. What Drives {SMiShing} Susceptibility? A {U.S}. Interview Study of How and Why Mobile Phone Users Judge Text Messages to be Real or Fake. 2024. 393–411. Retrieved August 12, 2024 from https://www.usenix.org/conference/soups2024/presentation/tabassum-sarah

[37] Daniel Timko, Daniel Hernandez Castillo, and Muhammad Lutfor Rahman. 2024. A Quantitative Study of SMS Phishing Detection. https://doi.org/10.48550/arXiv.2311.06911

[38] Melanie Volkamer, Karen Renaud, Benjamin Reinheimer, and Alexandra Kunz. 2017. User experiences of TORPEDO: TOoltip-poweRed Phishing Email DetectiOn. *Comput. Secur.* 71, (November 2017), 100–113. https://doi.org/10.1016/j.cose.2017.02.004

[39] Rundong Yang, Kangfeng Zheng, Bin Wu, Di Li, Zhe Wang, and Xiujuan Wang. 2022. Predicting User Susceptibility to Phishing Based on Multidimensional Features. *Comput. Intell. Neurosci.* 2022, 1 (2022), 7058972. https://doi.org/10.1155/2022/7058972

[40] 2018. SMS Phishing + Cardless ATM = Profit – Krebs on Security. Retrieved September 5, 2023 from https://krebsonsecurity.com/2018/11/sms-phishing-cardless-atm-profit/

[41] 2018. Cognitive Interviewing in Practice: Think-Aloud, Verbal Probing, and Other Techniques. . SAGE Publications, Inc., 42–65. https://doi.org/10.4135/9781412983655.n4

[42] 2019. Explore Data. *Federal Trade Commission*. Retrieved July 11, 2023 from https://www.ftc.gov/news-events/data-visualizations/explore-data

[43] 2022. Don't be a target: Phishing and smishing on the rise. *Sixteenth Air Force (Air Forces Cyber)*. Retrieved September 5, 2023 from https://www.16af.af.mil/Newsroom/Article/3138334/dont-be-a-target-phishing-and-smishing-on-the-rise/https%3A%2F%2Fwww.16af.af.mil%2FNewsroom%2FArticle-Display%2FArticle%2F3138334%2Fdont-be-a-target-phishing-and-smishing-on-the-rise%2F

[44] Number of mobile devices worldwide 2020-2025. *Statista*. Retrieved June 2, 2023 from https://www.statista.com/statistics/245501/multiple-mobile-device-ownership-worldwide/

[45] Smishing attacks up sevenfold in six months | Computer Weekly. *ComputerWeekly.com*. Retrieved March 29, 2023 from https://www.computerweekly.com/news/252506611/Smishing-attacks-up-sevenfold-in-six-months

[46] Avoid the Temptation of Smishing Scams. Retrieved June 6, 2023 from https://www.fcc.gov/avoid-temptation-smishing-scams

[47] 2023 Data Breach Investigations Report. *Verizon Business*. Retrieved September 1, 2023 from https://www.verizon.com/business/resources/reports/dbir/

[48] Smishing vs. Phishing vs. Vishing | HP® Tech Takes. Retrieved September 5, 2023 from https://www.hp.com/us-en/shop/tech-takes/smishing-vs-phishing-vs-vishing

[49] The 7 Latest Bank of America Scams You Need To Know. Retrieved December 10, 2023 from https://www.aura.com/learn/bank-of-america-scams

[50] New Report by Senator Warren: Zelle Facilitating Fraud, Based on Internal Data from Big Banks | U.S. Senator Elizabeth Warren of Massachusetts. Retrieved January 15, 2024 from https://www.warren.senate.gov/oversight/reports/new-report-by-senator-warren-zelle-facilitating-fraud-based-on-internal-data-from-big-banks

[51] Payments app Zelle begins refunds for imposter scams after Washington pressure | Reuters. Retrieved November 13, 2023 from https://www.reuters.com/technology/cybersecurity/payments-app-zelle-begins-refunds-imposter-scams-after-washington-pressure-2023-11-13/

[52] What is non-probability sampling? Definition and examples. *Qualtrics*. Retrieved January 30, 2025 from https://www.qualtrics.com/experience-management/research/non-probability-sampling/

# Appendix A: Text of Displayed Messages

## A.1 Study 1 messages – Diverse entities and scenarios

**R1**

From: 226-787

Message: Win a prize in our online scavenger hunt! BRU Information Security Office link:

https://www.bru.edu/iso/aware/ncsam/hunt/bonus.

**F1:**

From: +1 (310) 565-0315

Message: Walmart

Take our online survey and win a gift card! Log into your account at:

https://walm.me/8HHTz563AhouSf.

**R2:**

From: 721-66

Message: Chase Fraud: Did you use card 8201 at SHELL/SHELL on 06/13? Reply YES or NO or go to: https://www.chase.com/credit-cards/mobile/report-transaction. Rates may apply.

**F2:**

From: bankofamericahelp@customerdesk.com

Message: BOFA MSG: We locked your Debit Card to prevent unauthorized payment.To unlock it confirm your identity - rebrand.ly/xlk8w.

**R3:**

From: 226-787

Message: Baton Rouge U. Alumni exclusive free access - RSVP by 06/07 for our 06/21 Career Fair and upload your resume: https://www.bru.edu/career/resources/jobfair/.

**F3:**

From: +1 (540) 835-9232

Message: Amazon: Hiring NOW for

High-paid local expansion - reply YES to opt in & start background check at e3fmr.info/onAY235xU7.

**R4:**

From: +1 888-495-0312

Message: No obligation bankruptcy consultation - Apply online now at https://bankruptcylawyerfinder.com/freeconsultation.htm

**F4:**

From: +1 (877) 414-1179

Message: No obligation FREE security scan - check your phone for malware , link: http://mobile-validation-security.com/freescan.htm.exe

**R5:**

From: Ginger [image of a G in the profile icon]

Message: So glad u could join us Sat for M's bday! Here's my post, lmk what u think: https://www.tiktok.com/@mom28.edits20/video/2323g2323523

**F5:**

From: johns1985@gml.com

Message: Someone uploaded a video featuring your face http://128.3.72.234/nudepics2013.jpg.exe ;

**R6:**

From: 33339

Message: AMBER Alert

Amber Alert Vehicle TAN Honda Civic Tag Texas JDJ234. Call 911 https://amberalert.texas.gov

**F6:**

From: +1 (202) 609-0181

Message: WARNING:IRS audit pending - log in to dispute or appeal irS72.me/audit.htm on urgent basis otherwise your bank accounts, property, payments to be frozen by government.

**R7:**

From: 262966

Message: Re: Your Amazon Order (#103-0607555-6895008) - Problem with shipping may lead to delays. Confirm at www.amazon.com/help/confirmation.

**F7:**

From: Faceb0ok Security

Message: Facebook Alert: Your account has violated our policies! To avoid suspension visit bit.ly/2ucHM and reconfirm it. Facebook Security Team

## A.2 Study 2 messages – Bank of America entity, Zelle scenario

**Real1:**

From: 80814

Message: You sent a Zelle payment to TRAVIS. Amount: $75. Date: 02/02/24. ID: BOAAA0IJI53i. For more information, log onto bankofamerica.com/zelle.

**Real2:**

From: 80814

Message: You sent a Zelle payment to TRAVIS. Amount: $75. Date: 02/02/24. ID: BOAAA0IJI53i. For more information, call 1-800- 482-1000.

**Real3a:**

From: 80814

Message: <#>BofA: You sent a Zelle payment to TRAVIS. Amount: $75. Date: 02/02/24. ID: BOAAA0IJI53i. For more information, reply INFO.

**Real3b:**

From: 80814

Message: <#>BofA: You sent a Zelle payment to TRAVIS. Amount: $75. Date: 02/02/24. ID: BOAAA0IJI53i. For more information, call the number listed on the back of your BofA card.

**Fake1a:**

From: bankofamericahelp@customerdesk.com

Message: You sent a Zelle payment to TRAVIS. Amount: $75. Date: 02/02/24. ID: BOAAA0IJI53i. For more information, log onto https://boaonline.s3.amazonaws.com/tedmSb.html?Hked.

**Fake1b:**

From: (888) 262-5692

Message: You sent a Zelle payment to TRAVIS. Amount: $75. Date: 02/02/24. ID: BOAAA0IJI53i. For more information, log onto https://boaonline.s3.amazonaws.com/tedmSb.html?Hked

**Fake1c:**

From: 29450

Message: You sent a Zelle payment to TRAVIS. Amount: $75. Date: 02/02/24. ID: BOAAA0IJI53i. For more information, log onto https://boaonline.s3.amazonaws.com/tedmSb.html?Hked

**Fake2a:**

From: bankofamericahelp@customerdesk.com

Message: You sent a Zelle payment to TRAVIS. Amount: $75. Date: 02/02/24. ID: BOAAA0IJI53i. For more information, call us at 201-416-7037

**Fake2b:**

From: (888) 262-5692

Message: You sent a Zelle payment to TRAVIS. Amount: $75. Date: 02/02/24. ID: BOAAA0IJI53i. For more information, call us at 201-416-7037

**Fake2c:**

From: 29450

Message: You sent a Zelle payment to TRAVIS. Amount: $75. Date: 02/02/24. ID: BOAAA0IJI53i. For more information, call us at 201-416-7037

**Fake3a:**

From: bankofamericahelp@customerdesk.com

Message: <#>BofA: You sent a Zelle payment to TRAVIS. Amount: $75. Date: 02/02/24. ID: BOAAA0IJI53i. For more information, reply INFO or 1 To Unsubscribe Texts

**Fake3b:**

From: (888) 262-5692

Message: <#>BofA: You sent a Zelle payment to TRAVIS. Amount: $75. Date: 02/02/24. ID: BOAAA0IJI53i. For more information, reply INFO or 1 To Unsubscribe Texts

**Fake3c:**

From: 29450

Message: <#>BofA: You sent a Zelle payment to TRAVIS. Amount: $75. Date: 02/02/24. ID: BOAAA0IJI53i. For more information, reply INFO or 1 To Unsubscribe Texts

## Appendix B: Testing a Role-Play Scenario

Downs, Holbrook, and Cranor [13] established the method of "embedded role play" for simulated phishing research. For their 2007 paper, participants were asked to respond to simulated emails and web sites as "Pat Jones," a researcher at the fictional company Cognix. Participants reported taking the role-play seriously. Many reported responding in ways consistent with how they themselves would judge the credibility of emails and web sites, but applying their own knowledge and experience to evaluate purported content from real-life companies such as PayPal or NASA that they themselves did not hold accounts with.

As part of Study 1, we investigated whether the "Pat Jones" role-play developed by Downs et al. continues to show results similar to those from questions answered by participants who were not asked to role-play. We evenly randomized all participants into two survey conditions: judging the SMS text messages as either "yourself" (described as whether you, the participant, had received the message on your phone) or as "Pat Jones." For our role-play, we adapted the text from the online survey in Sheng et al. [34], in which "Jones" was described as a staff member of Baton Rouge University who has many accounts and whose job makes it important to not fall for fraudulent text messages and to respond promptly to legitimate text messages. This enabled us to collect data to assess whether the embedded role-play method remains valid for simulation studies such as ours and to statistically control for its influence in our regression analyses, if needed.

A regression revealed that no significant variance in CORRECT could be explained by whether the participants had answered the questions as "yourself" or as "Pat Jones" (Adj. $R^2$=-.001, $F$(1,1004)=.104, $p$=.747). We concluded from this that there was no need to control for condition assignment in inferential analyses of other data. Further, for those who were assigned to the "Pat Jones" role-play condition, no significant variance in CORRECT could be explained by self-reported knowledge of having an account with the named entity (Adj. $R^2$=-.002, $F$(1,492)=.166, $p$=.684). This seems to support the argument from Downs et al. [13] that participants in simulation studies will take the role play seriously and answer not as themselves but as the fictional character whom they are instructed to answer as.

## Appendix C: Study 1 Open-Ended Responses

| | How to respond | Why that response |
|---|---|---|
| R1 | Contact school tech team to see if this is a true message sent by them.<br>report to phone service provider<br>Contact BRU<br>block number<br>Check for a web site for BRU.edu<br>Links can be manipulated. <a href="xxx.html">aaa.html</a> | No<br>If it is a scam, my personal data will be compromised.<br>Caution<br>I would have known about this in advance. I wouldn't be texted out of the blue.<br>I usually don't click prize scams in SMS messages. Too much of a security issue, unless if it's from the company/business's official number/account. |
| F1 | report to phone service provider<br>Action dependent if we had an account with Walmart<br>Ignore the text unless I felt that I wanted to take the survey... just like I am taking this one.<br>Call walmart<br>report to walmart | To not fall for the scam<br>I won't respond back.<br>Scam<br>the return web address is not walmarts web address. |
| R2 | Check my bank account<br>Look into my bank account and see if there was a transaction made that I did not do<br>Google for similar text messages<br>Call chase<br>Call the credit card company directly<br>Call company<br>I would call Chase to see if they sent the text<br>[C]all the bank<br>Go to my bank website directly without using a link to check and see if there are notifications on my account page. Also google to see if similar texts have been reported as scams.<br>I would contact a legitimate number from my info when opening my account to inquire about any fraudulent activeit on my accounty<br>depends on if I made a purchase<br>If I had a Chase card with those ending digits, I'd call their customer support directly to resolve the issue.<br>go directly to chase and see if that amount was charged.<br>If I actually have a Chase card and actually did use it for gas on that date, then I would respond to the msg. Otherwise, I would report it to Chase as potential fraud.<br>Log on to my Chase account and react to alert, there.<br>Call bank on phone number I know is correct<br>I would not reply to anything, but I would<br>call Chase bank to see if this is a scam<br>If I had a card with them I would call and check it out<br>Call chase<br>Telephone bank cust svc number to discuss<br>Phone Chase Bank to verify<br>Check my Chase account online<br>Forward the message to my bank. Banks don't put RATES MAY APPLY in the message. | I wouldnt<br>Knowing I'd probably get my information stolen<br>Safer than clicking on fraudulent link or responding to text<br>Don't like scams<br>I am proactive.<br>Prefer to work directly on my bank site, for all inquiries and transactions<br>Link begins with HTTPS<br>Think it's a scam<br>Delete with no response<br>I don't click links unless family. |
| F2 | would call them<br>check the status of my debit card on the providers website<br>I don't know<br>type it into a web browser in a separate device<br>Call the bank<br>Call the company<br>call my bank<br>call bank of america<br>Call my credit card company to verify this is true information<br>Sign into my account and see if it works.<br>Call bank to check if legitimate<br>call my bank<br>If I actually had a BOA account, I would contact them directly (not via a reply to the SMS) and verify the info. If it is legit, I would take appropriate action(s).<br>I would call bank of America to let them know I received this message<br>Contact the bank by phone to verify they sent the text.<br>I'd contact the number for customer service on the back of my card and check with a real person<br>call the bank directly<br>Forward it to whoever handles phishing<br>If I did have an account with them i would alternately verify that it was indeed blocked and and then remedy.<br>Telephone my bank cust svc number to ask if they sent msg<br>call bank of America | leave it alone<br>I don't know<br>I won't respond back nor click on the link.<br>Just doesn't look real<br>probable scam<br>the senders email address is not consistant with a address from BOAt<br>I want to see if this is a scam<br>Don't know how to do computer stuff on phone. |
| R3 | (no Other responses) | scam<br>think its a scam and don't want to risk it.<br>Not relevant to me in any way<br>scam |
| F3 | Also, block the number<br>Delete<br>Would report to amazon, who does follow up on scammers using their brand name. | No<br>possible scam<br>Dont need/want a job<br>Red flags in this. |
| R4 | Call company<br>Spam at least, to dangerous to click link for any reason DELETE | I seek out what I want and don't respond to things I didn't initiate<br>I won't respond back.<br>scam<br>am not near bankruptcy<br>I don't believe that they would send a message |

| | How to respond | Why that response |
|---|---|---|
| | | Unknown individual sending me an interested text<br>SPAM |
| F4 | block number | I don't click on random links sent to me regardless of their claim<br>to good to be true usually means............<br>Report scammers to provider<br>Scam<br>Caution<br>Looks like a scam spam at the least, dangerous to click link DELETE |
| R5 | Text ginger and ask if she sent it or say something she would respond to to verify its her<br>If I really went to a birthday party I'd click the link<br>Message back to talk to someone to make sure it is reL<br>Select the link only if I had previous knowledge of what the message was about<br>I would try to remember if there is a Ginger among my friends<br>respond if i know ginger<br>Call ginger yo see if she really sent the text<br>text the person who had the party to see if they sent it to me.<br>Don't know how to read texts.<br>If we knew Ginger, and M and attended the birthday party, we would respond. | No<br>Scammers need to be reported<br>Don't read or answer texts. |
| F5 | (no Other responses) | scam<br>NOT A LEGIT EMAIL ADDRESS<br>Scam. I see this a lot |
| R6 | wait for the next sms message.<br>Try remember text<br>Read it, possibly save it, or delete it after<br>Keep a look out for the vehicle<br>nothing unless I have info then I would call 911 to give the info I had<br>Keep a lookout for the vehicle.<br>Turn on the radio or the news<br>Review it and be on lookout for said vehicle. Later on would delete.<br>Be on the lookout for that license #<br>Nothing much; would read, look up news, and clear emergency alerts in my storage.<br>My notification repeats until disabled.<br>Ignore it unless I spotted a Tan Honda Civic with the license plate listed in the text.<br>Call Law enforcement to see if it is valid. | moral obligation to keep children safe<br>An amber alert wouldn't have a link to push<br>I would believe it but wouldn't click anything<br>there's no description of the kidnapped person in the message at all, so it's most likely a fraudulent text<br>keep alert from that txt, but nothing further of action to be done.<br>Right thing to do<br>Amber alerts don't have links<br>Has a .gov address so will be on alert for car but won't click the link<br>Important to be on |

| | How to respond | Why that response |
|---|---|---|
| | | lookout for missing children.<br>would mostly like not know anything to reply to message<br>Didn't delete incase I would happen to see something.<br>No need to respond unless I see the vehicle in question.<br>no knowledge of it<br>Amber alerts come as a notification, not as a text<br>Keep myself safe while being a good citizen<br>Skepticism<br>I generally don't click on links unless sent from personal contact<br>By calling 911 it would validate if the message was a scam<br>I would save it to call in case I saw the vehicle in question.<br>Not living near that state<br>look out for a Honda with Texas tags.<br>Amber alerts do not come over sms texts |
| F6 | copy it into a browser on a separate device as an anonymous<br>report to phone service proivider<br>look up the phone number<br>the IRS only send a notice in the mail | To be safe<br>IRS communicates by certified mail.<br>SPAM<br>Because the government wouldn't send a text like that<br>I don't respond to spam<br>I feel like all the messages are scam nowadays<br>The IRS only notify you by mail<br>It's a scam<br>Caution<br>Highly suspect message<br>IRS only contacts people thru the USPS<br>IRS would not send a text<br>It's wouldn't send a message.<br>I know the IRS would never send me a text message |
| R7 | log on into Amazon's account<br>Forward to Amazon<br>Go to my Amazon account<br>Block sender | it will be accurate<br>My shipping's contact info, where "262966" is in image, |

| | How to respond | Why that response |
|---|---|---|
| | I'd just get on my Amazon app<br>Go to my Amazon account and look for notifications<br>Check my account<br>Go to the Amazon web site directly to see what's up<br>Block number<br>Use my internet to contact Amazon to see if the message is real or a fruad.<br>contact Amazon directly, not use the text<br>Report to Amazon<br>call Amazon to verify<br>Check email or go to Amazon site<br>Need to tell me where my package is to show me if there is a problem. | is saved from initial app verification code sent. Any other's are ignored. I ignore things like this and go to app or website. And Gmail; I check sender's address there<br>it would depend on if I had ordered anything from Amazon<br>Needs to be from the official carrier to be legitimate. |

| | How to respond | Why that response |
|---|---|---|
| F7 | Check to see if my FB account is actually suspended and if it hasn't, I would report the sender.<br>I would check my fb, but not from that link<br>Sign into Facebook to see if there are any problems. If necessary send a message within Facebook to ask if there is a problem.<br>check my facebook account | Safer to verify the information in another way rather than responding to possible fraudulent texts.<br>Hate scams and spam<br>Scam<br>I don't use social Media<br>Scam |

## Appendix D: Survey Protocols

**Study 1 survey: https://drive.google.com/file/d/1RlJHlgQ4ri-KwbafWuJr8Y_Z58herNxGX/view?usp=sharing**

**Study 2 survey:**
**https://drive.google.com/file/d/1W6gX1vBzVmfHIKfWCvx3ehk2SPJGPzJz/view?usp=sharing**

### Consent Form – Study Summary Section

You are invited to participate in a research study. Participation in this research study is voluntary. The information provided is to help you decide whether or not to participate. If you have any questions, please ask.

Important Information You Need to Know

• The purpose of this study is to document how different types of people interpret various mobile phone text messages.

• You must be a U.S. mobile phone user age 18 or older to participate in this study.

• You will be asked to answer a number of questions in an online survey. Most are multiple choice.

• If you choose to participate, it will require [x] minutes of your time.

• Risks or discomforts from this research are rare, but may include mild annoyance or frustration with the questions being asked, or discomfort with personal reflections about past experiences.

• While you will receive no direct benefit, the knowledge gained will be of value for improving people's experiences while using mobile phones and receiving text messages.

• If you choose not to participate, you may exit the survey.

• If you participate and complete the study, you will receive compensation in a form that is consistent with your agreements with the survey panel provider.

Please read this form and ask any questions you may have before you decide whether to participate in this study.

### Directions

For the following three screens, please rate the following:

• How likely is it that the given SMS text message is a legitimate message or a scam message,

• What would you be likely to do in response, and why.

In each case, make your choice in terms of how you feel right now, not what you have felt in the past or would like to feel.

### Message rating block

Remember to answer each question as yourself.

<code to insert image>

***Study 1 rating options***

Indicate the degree to which you yourself think this message is legitimate or a scam:

o Almost certain to be a scam
o More likely to be a scam than to be legitimate
o Equally likely to be either legitimate or a scam
o More likely to be legitimate than to be a scam
o Almost certain to be legitimate

***Study 2 rating options***

Do you think that this message is real or a scam?

o Real - this looks like a legitimate message
o A scam - this looks like a fraudulent message

How confident are you in your answer above?

o Not at all confident
o A little confident
o Moderately confident
o Very confident

o Absolutely confident

If you yourself received this message on your mobile phone, what would you do in response? Select all that apply.

▢ Reply by SMS text message to provide information
▢ Reply by SMS text message with STOP, BLOCK, or other codes intended to stop receiving messages from the sender
▢ Report the message using your device options (such as clicking Block This Caller or Report Junk)
▢ Click on the link in the SMS text message
▢ Copy and paste the link into your phone's web browser
▢ Type the link into your phone's web browser
▢ Forward the SMS text message to someone else
▢ Delete the SMS text message
▢ Keep, save, or archive the SMS text message
▢ Ignore the SMS text message
▢ Other ________________________________

Why would you respond this way? Select all that apply.

▢ Sense of urgency
▢ Curiosity
▢ Trust in the sender
▢ Trust in the link URL (if included)
▢ Seeking a good outcome for myself
▢ Seeking to avoid a bad outcome for myself
▢ Lack of interest in the message
▢ Other ________________________________

***Study 1 follow-up***

To the best of your knowledge, do you yourself have an account with the entity mentioned in the text message?

o Yes
o No
o Not sure

***Study 2 follow-up***

Are you, in fact, a customer of Bank of America?

o Yes
o No
o Not sure

Are you, in fact, a user of the Zelle mobile payments service?

o Yes
o No
o Not sure